\documentclass[aps,prb,preprint,superscriptaddress]{revtex4-1}

\usepackage{amsmath,amssymb,amstext}
\usepackage{mathrsfs}
\usepackage{dsfont}
\usepackage{graphicx}
\usepackage[all]{xy}
\usepackage{color}

\newcommand{\cB}{{\mathcal B}}
\newcommand{\cD}{{\mathcal D}}
\newcommand{\cL}{{\mathcal L}}
\newcommand{\cK}{{\mathcal K}}

\newcommand{\cH}{{\mathcal H}}
\newcommand{\cO}{{\mathcal O}}

\newcommand{\cR}{{\mathcal R}}
\newcommand{\cS}{{\mathcal S}}
\newcommand{\cT}{{\mathcal T}}
\newcommand{\cW}{{\mathcal W}}

\newcommand{\cV}{{\mathcal V}}
\newcommand{\cU}{{\mathcal U}}
\newcommand{\un}{{\mathds 1}}
\newcommand{\cQ}{{\mathcal Q}}
\newcommand{\RR}{{\mathbb R}}

\newcommand{\CC}{{\mathbb C}}

\newcommand{\mH}{{\mathscr H}}

\newcommand{\mD}{{\mathscr D}}
\newcommand{\mO}{{\mathscr O}}

\newcommand{\eps}{\varepsilon}

\newcommand{\matel}[3]{{\left\langle \vphantom{#1 #2 #3} #1 \,\right\vert
\left.
 \hspace{-0.15em} \vphantom{#1 #2 #3} #2 \,\right\vert \left.
 \hspace{-0.15em} \vphantom{#1 #2 #3} #3\right\rangle}}

\newcommand{\pa}{\partial}
\newcommand{\tr}[1]{\mathrm{tr}\left(#1\right)}

\newcommand{\trB}[1]{\mathrm{tr}_B\left(#1\right)}

\newcommand{\be}{\begin{equation}}
\newcommand{\ee}{\end{equation}}
\newcommand{\bea}{\begin{eqnarray}}
\newcommand{\eea}{\end{eqnarray}}

\newcommand{\dint}[1]{\mathrm{d} #1}
\newcommand{\sinto}[1]{\int\mathrm{d} #1 }
\newcommand{\dinto}[2]{\int\mathrm{d} #1 \mathrm{d} #2 }

\renewcommand{\v}[1]{\ensuremath{\vec{#1}}}

\begin{document}

\title{Derivation of a linear collision operator for the spinorial Wigner equation and its
semiclassical limit}

\author{Benjamin A. Stickler}
 \email[]{benjamin.stickler@uni-graz.at}
 \affiliation{Institute of Physics, Karl-Franzens Universit\"at Graz, A-8010 Graz, Austria}
\author{Stefan Possanner}
 \email[]{s.possanner@gmail.com}
 \affiliation{Institut de Math\'ematiques de Toulouse, Universit\'e Paul Sabatier, Toulouse, France}

\begin{abstract}
We systematically derive a linear quantum collision operator for the spinorial Wigner transport equation from the dynamics of a composite quantum system. For suitable two particle interaction potentials, the particular matrix form of the collision operator describes spin decoherence or even spin depolarization as well as relaxation towards a certain momentum distribution in the long time limit. It is demonstrated that in the semiclassical limit the spinorial Wigner equation gives rise to several semiclassical spin-transport models. As an example, we derive the Bloch equations as well as the spinorial Boltzmann equation, which in turn gives rise to spin drift-diffusion models which are increasingly used to describe spin-polarized transport in spintronic devices. The presented derivation allows to systematically incorporate Born-Markov as well as quantum corrections into these models.
\end{abstract}

\maketitle

\section{Introduction} \label{sec:intro}

The modeling of transport phenomena in electronic devices is one of the major challenges in modern solid state physics. While in most physical applications a full quantum mechanical treatment by means of the Schr\"{o}dinger equation or the von Neumann equation is far too complex, it is a beneficial and legitimate approach to employ effective models. Well-known and prosperous examples are the drift-diffusion equations to treat systems in local thermal equilibrium and the Boltzmann equation (BE) to capture non-equilibrium phenomena.

It is a major challenge to clarify the simplifactions and approximations posed in a microscopic theory which lead to such an effective model. We state spintronics as an example\cite{fabian1,fabian2} where spin-drift-diffusion equations proved to be a powerful tool for describing spin-polarized  transport\cite{zhang02} and spin-transfer torques \cite{garcia06,possanner10} in magnetic mulitlayers. It has been demonstrated that these equations can be derived from a spinorial BE \cite{simanek,raymond,MeClaudia2011}. Hence, the missing link in a systematic, qualitative understanding is the derivation of a spinorial BE starting from a full quantum mechanical treatment. This is the main goal of the present study.

Let us briefly discuss some well-established results in order to position the present work in an appropriate context: The generic form of the scalar BE is
\be \label{BE}
\pa_t f - \{h,f\}_{x,\eta} = C(f)\,,
\ee
where $f(x,\eta,t)$, $f \geq 0$, is a probability distribution on the $2d$-dimensional phase space $\RR_x^d\times\RR^d_\eta$, $h(x,\eta)$ stands for the energy of a non-interacting particle and $\{h,f\}_{x,\eta}=\nabla_x h\cdot \nabla_\eta f - \nabla_\eta h \cdot \nabla_x f$ denotes the Poisson bracket with respect to the position coordinate $x$ and the momentum coordinate $\eta$. The collision operator $C$ on the right-hand-side (rhs) of \eqref{BE} models short range interactions between particles or with obstacles, e.g. impurity centers or phonons in case of electronic transport in semiconductor devices. $C$ is usually an integral operator and, moreover, non-linear in case that it describes interactions between identical particles or accounts for quantum statistics. Eq. \eqref{BE} is referred to as the semiclassical BE since microscopic properties like the electronic bandstructure and quantum scattering rates can be described in terms of $h(x,\eta)$ and $C(f)$, respectively.

The incorporation of further quantum phenomena like coherence and entanglement \cite{nielsen, awschalom, zeilinger} creates a need for either quantum corrections to the BE or quantum versions thereof, called quantum Boltzmann equations \cite{jauho, spicka, hornberger}.  Moreover, the recent emergence of spintronics \cite{fert} raised the question of how to describe scattering of spin-coherent electron states in magnetic multilayers or domain walls by means of a kinetic equation \cite{thiaville, simanek, xiao, culcer, zhang}. In the spin-coherent regime, the BE \eqref{BE} is replaced by
\be \label{BE2}
\pa_t F - \{h\mathds{1},F\}_{x,\eta} + i[\Omega,F]  = Q(F)\,,
\ee
where $F(x,\eta,t)$, $\Omega(x,\eta,t)$ are hermitian $2\times2$ matrices defined on the phase space and $[\Omega,F]=\Omega F - F \Omega$ denotes the commutator. $F$ is the distribution matrix, the eigenvalues of which give the scalar distribution functions of the two spin species. The term $\Omega$ is an exchange field that mixes the two spin distributions. Equation \eqref{BE2} is referred to as the spinorial or matrix Boltzmann equation (SBE)\cite{MeClaudia2011,raymond}. Possanner and Negulescu\cite{MeClaudia2011} studied linear collision operators $Q$ which feature spin-dependent scattering rates, for example
\be \label{QF}
Q(F)(\eta) = \int \dint{\eta'} \left( S'^{1/2}F(\eta')S'^{1/2} - \frac{1}{2}SF(\eta) - \frac{1}{2}F(\eta)S \right)\,,
\ee
Here, $S = S(\eta,\eta')$ is a strictly positive, hermitian $2\times2$ matrix, whose eigenvalues denote the scattering rates from $\eta$ to $\eta'$ for the two spin species \cite{fertscattering, viret} and $S'=S(\eta',\eta)$. The left-hand-side (lhs) of Eq. \eqref{BE2} has been derived on a rigorous basis by Hajj\cite{raymond}. A derivation of \eqref{QF}, which is able to relate the scattering matrices $S$ to a microscopic Hamiltonian will be accomplished in the course of this work.

The derivation of QBEs or the SBE starts at the microscopic level by defining a suitable model Hamiltonian. Then, the natural framework to pass from the quantum to the kinetic level is the Wigner-Weyl formalism of quantum mechanics \cite{zachos, gerard}. There exists a plethora of results regarding this passage for the scalar (spin-less) case, some of them we shall briefly mention here (for further information the reader is urged to view the references in the articles cited below). We remark that for the case that the eigenvalues of $S$ are identical, performing the trace in Eq. \eqref{BE2} leads the scalar BE \eqref{BE}. QBEs have been obtained in the framework of generalized Kadanoff-Baym non-equilibrium Green's functions \cite{jauho, spicka, jauhobook} and by a monitoring technique \cite{hornberger}. On the rigorous level, the linear BE has been obtained from the single particle Schr\"odinger equation with a Gaussian random potential in the weak-coupling \cite{erdos2000, spohn77} and in the low-density 
limit \cite{erdos2005}, respectively. The non-linear BE was derived by starting from the many-body Schr\"odinger equation 
with weak pair interaction potential and by studying the quantum version of the BBGKY-hierarchy.\cite{pulvirenti} 

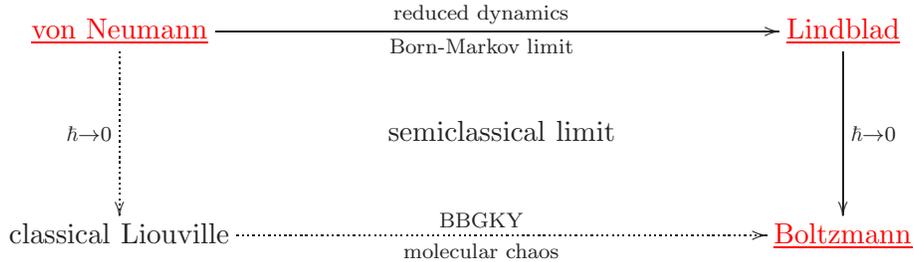
\begin{figure} 
$$
\xymatrix{
{\textcolor{red}{\underline{\textrm{von Neumann}}}} \ar@{.>}[dd]_{\hbar\to 0} \ar[rrrr]^{\textrm{reduced dynamics}}_{{\textrm{Born-Markov limit}}} & & & & 
{\textcolor{red}{\underline{\textrm{Lindblad}}}} \ar[dd]^{\hbar\to 0} \\
& &\textrm{semiclassical limit} & &  \\
\textrm{classical Liouville} \ar@{.>}[rrrr]^{\textrm{BBGKY}}_{\textrm{molecular chaos}} & & & &  {\textcolor{red}{\underline{\textrm{Boltzmann}}} }}
$$
\caption{(Color online)Schematic illustration of the strategies to pass from the quantum level to a semiclassical BE. The path employed in the present work is indicated by  solid arrows while an alternative approach is indicated by dotted arrows.} \label{fig:limit}
\end{figure}

In this work we apply a different strategy for passing to the kinetic level. Our starting point for the semiclassical analysis will be a master equation of the Lindblad form \cite{lindblad} describing a single quantum particle in contact with its environment\cite{breuer,rivas}. Semiclassical limits of Lindblad type master equations have already been considered.\cite{almeida} In particular, we shall start from the hierarchy of master equations derived by Possanner and Stickler\cite{possanner12}. Master equations of Lindblad form describe the quantum evolution in terms of a semigroup law (quantum dynamical semigroups) \cite{kossakowski}, just as the BE does for the classical evolution on the kinetic level. Therefore, by starting the semiclassical analysis from the Lindblad equation instead of the von Neumann equation, the passage from the quantum to the kinetic level has been decomposed into two stages as sketched in Fig. \ref{fig:limit}:
\begin{enumerate}
\item[(1)] In the quantum regime, one performs a Markovian limit that leads to dynamics described in terms of a quantum dynamical semigroup (Born-Markov limit\cite{possanner12}),
\item[(2)] in the Markovian regime, one performs the semiclassical limit (scaled $\hbar\to 0$) in order to obtain the BE.
\end{enumerate}
Corrections to the BE arise at each of the two stages. One obtains non-Markovian corrections at the first stage and quantum corrections in ascending powers of $\hbar$ (scaled) at the second stage. This paper deals solely with the second stage, while the first has been accomplished by Possanner and Stickler\cite{possanner12}.

This paper is organized as follows: in Sec. \ref{sec:notation} we shall agree on some notations and specify the physical system under investigation. In Sec. \ref{sec:intkern} we explicitly evaluate the integral kernel of the dissipator, in Sec. \ref{sec:wigner} we introduce the Wigner transform of the state operator and transform the whole master equation into the Wigner representation. Here we derive a Wigner equation equipped with a linear collision operator which features momentum relaxation as well as spin decoherence. The details of the derivation are explicated in App. \ref{app:fourier} - App. \ref{app:dissicomm}. In Sec. \ref{sec:coll} we discuss the quantum collision operator and, finally, in Sec. \ref{sec:scalenlimit} we introduce the semiclassical scaling and define the different semiclassical scenarios which we will regard in this work. Moreover, we draw the semiclassical limit for these scenarios and, thus, derive the spinorial BE with a collision operator of the 
form \eqref{QF} as well as the Bloch equations. Conclusions are drawn in Sec. \ref{sec:summary}.

\section{Notations and modeling} \label{sec:notation}

We consider the evolution of a single quantum particle with two spin degrees of freedom (spin $1 / 2$-particle), henceforth called the {\em 'system'}. This particle interacts with the {\em 'environment'} $B$ which in turn is composed of $N$ identical spin $1/2$-particles.

The dynamics of the system's particle are governed by a master equation of the Lindblad form\cite{possanner12},
\be \label{eq:hier}
\partial_t \hat{\rho} = -\frac{i}{\hbar} [\hat{H}^{mf}_0, \hat{\rho} ] + \cD( \hat{\rho} ).
\ee
Here, $\hat \rho$ is the density matrix defined on a one-particle Hilbert space $\cH$, $\hbar$ denotes the Planck constant, $\hat{H}^{mf}_0$ and $\cD$ stand for the system Hamiltonian and the dissipator, respectively. The Hamilton operator $\hat{H}^{mf}_0$ acting in $\cH$ reads $\hat{H}^{mf}_0 = \hat{H}_0 + \hat{H}_{mf}$, where $\hat{H}_0$ is a one-particle Hamiltonian and one defines the mean-field operator
\be \label{Hmf}
\hat{H}_{mf} = \trB{\hat{H}_I \hat{\un} \otimes \hat{\chi}_B}\,.
\ee
Here, $\hat{H}_I$ is an operator acting in the composite Hilbert space $\cH \otimes \cH_B$, which describes the interaction between the system's particle and the environment (or bath), $\hat \un$ acting on $\cH$ is the unity operator in $\cH$ and $\hat{\chi}_B$ is a predefined equilibrium density matrix on $\cH_B$. The operation $\trB{\cdot}$ stands for taking the trace over the degrees of freedom of the bath. Moreover, the action of the dissipator $\cD$ is defined by 
\be \label{eq:dissi}
\cD(\cdot) := - \frac{\tau_0}{\hbar^2} \trB{[\hat{H}^{mf}_I, [\hat{H}^{mf}_I, \cdot \otimes \hat \chi_B]]},
\ee
where $\tau_0$ denotes the characteristic timescale\footnote{Denoting by $\tau_0$ the characteristic timescale of the system corresponding to the Hamiltonian $\hat H^{mf}_0$, the characteristic energy is defined via $\epsilon_0 \tau_0 = \hbar$.} of the system's dynamics $\hat H^{mf}_0$ and $\hat H^{mf}_I = \hat H_I - \hat H_{mf} \otimes \hat \un_B$. Please note that in writing \eqref{eq:dissi} we assume that the Hamiltonian $\hat H^{mf}_I$ is associated with the same characteristic energy $\epsilon_0$ as $\hat H^{mf}_0$, i.e. we rescaled $\hat H^{mf}_I \to \hat H^{mf}_I \epsilon_0 / \epsilon_I$ where $\epsilon_I$ denotes the characteristic mean-field corrected interaction energ \cite{possanner12}. Equations \eqref{eq:hier}-\eqref{eq:dissi} are valid for very fast relaxation of bath states towards $\hat \chi_B$ and $\tau_I / \tau_0 \ll 1$. It has to be emphasized that due to the assumptions incorporated in the derivation of Eq. \eqref{eq:hier} we restrict our discussion to a case 
in which the system's particle is distinguishable from the particles constituting the environment. Furthermore, we note that under certain premises, Eq. \eqref{eq:hier} may account for the dynamics of the system's particle towards a unique equilibrium state.\cite{esposito11}

Let us briefly comment on the physical picture employed: We assume that Eq. \eqref{eq:hier} provides a proper description of the quantum dynamics of the system's particle in contact with its environment. It is the aim of this work to draw the semiclassical limit of Eq. \eqref{eq:hier}, i.e. to regard the dynamics of the system's particle in a regime in which quantum effects cease to be observable. This goal is achieved in three steps: in a first step we shall rewrite Eq. \eqref{eq:hier} as an equation for the integral kernel $\rho(x,x',t) = \matel{x}{\hat \rho(t)}{x'}$ in position space, in a second step we shall derive the Wigner representation of Eq. \eqref{eq:hier} and, finally, in a third step we shall draw the semiclassical limit of Eq. \eqref{eq:hier}.

However, we need to clarify some notations first. In what follows we shall denote the position and momentum coordinates of the system's particle by $x \in\RR^d_x$ and $\eta \in \RR_\eta^d$, respectively, and the position coordinate of the $n$-th particle in the environment $B$ by $z_n \in \RR_{z_n}^d$. Moreover, we introduce the short-hand notation $Z = (z_1,z_2, \ldots, z_N) \in \RR_Z^{Nd}$. For multiple integrals we use the abbreviations
\begin{subequations}
\begin{align}
 \sinto{Z} &= \sinto{z_1}\sinto{z_2}\ldots\sinto{z_N}\,,\\
 \sinto{Z_n} &= \sinto{z_1}\ldots\sinto{z_{n-1}}\sinto{z_{n+1}}\ldots\sinto{z_N}\,,\\
 \sinto{Z_{nm}} &= \sinto{z_1}\ldots\sinto{z_{n-1}}\sinto{z_{n+1}}\ldots\sinto{z_{m-1}}\sinto{z_{m+1}}\ldots\sinto{z_N}\,.
\end{align}
\end{subequations}
The spin degrees of freedom of the system will be labeled by roman lower case letters, such as $i,j \in (1,2)$, while the spin degrees of freedom of the $n$-th particle in $B$ are labeled by Greek letters, i.e. $\alpha_n, \beta_n \in (1,2)$. In what follows we shall use the multi-index notations
\begin{subequations}
\begin{align}
\{\alpha \} &= (\alpha_1, \alpha_2,\ldots, \alpha_N)\,, \label{mui1}\\
\{\alpha \}_{n} &= (\alpha_1, \ldots, \alpha_{n-1}, \alpha_{n+1},\ldots, \alpha_N)\,,\\
\{\alpha \}_{nm} &= (\alpha_1, \ldots, \alpha_{n-1}, \alpha_{n+1},\ldots, \alpha_{m-1}, \alpha_{m+1}, \ldots \alpha_N)\,,\\
\{\alpha,\beta_n \} &= (\alpha_1, \ldots, \alpha_{n-1}, \beta_n, \alpha_{n+1},\ldots, \alpha_N)\,. \label{mui4}
\end{align}
\end{subequations}
In particular, for the multi-indices \eqref{mui1}-\eqref{mui4} we employ the abbreviation $\sum_{ \{ \alpha \} } \equiv \sum_{\alpha_1 \alpha_2 \cdots \alpha_N}$ and write the matrix elements of a $2^N\times 2^N$ matrix $\Gamma$ as $\Gamma^{\{\alpha\},\{\beta\}} \equiv \Gamma_{ \alpha_1\alpha_2\cdots\alpha_N,\beta_1 \beta_2 \cdots \beta_N}$. 

In what follows the set of hermitian $2\times 2$ matrices is termed $\mH_2(\CC)$ and we shall use the notation $\v{\sigma}=(\sigma_1,\sigma_2,\sigma_3)$, where
\be
\sigma_{1}=\begin{pmatrix}
             0&\:\:1\\1&\:\:0
            \end{pmatrix}\,,
\qquad\sigma_{2}=\begin{pmatrix}
                   0&-i\\i&0
                  \end{pmatrix}\,,
\qquad\sigma_{3}=\begin{pmatrix}
                   1&0\\0&-1
                  \end{pmatrix}\,,
\ee
are the three Pauli matrices. Hence, any matrix $G\in\mH_2(\CC)$ can be written in the Pauli basis $(\mathds{1},\v{\sigma})$ with coefficients $g_0 \in \RR$ and $\v{g}=(g_1,g_2,g_3) \in \RR^3$, respectively,
\begin{subequations} \label{eq:paulibase}
\begin{gather}
 G = g_0 \mathds{1} + \v{g}\cdot\v{\sigma}\,, \\
 g_0 = \frac{1}{2}\tr{G}\,,\qquad \v{g} = \frac{1}{2}\tr{\v{\sigma}G}\,,
\end{gather}
\end{subequations}
where $\mathds{1}$ is the $2 \times 2$ unity matrix. The eigenvalues of $G$ can be expressed as $g_0 \pm \vert \vec g \vert$, hence we call $\vert \vec g \vert$ the spin polarization of $G$. Moreover, we refer to $\vec g$ as the spin (spinorial) part or directional spin polarization and to $\vec g / \vert \vec g \vert$ as the direction of spin polarization of $G$.\footnote{We note that in the case of more than two spin degrees of freedom one may express all quantities in a form analogous to Eq. \unexpanded{\eqref{eq:paulibase}} by employing the generalized Gell-Mann matrices\unexpanded{\cite{bertlmann08}}.} Moreover, we remark the following property: let $A,B \in \mH_2(\CC)$, then with the help of the decomposition \eqref{eq:paulibase} and the properties of the Pauli matrices we have
\be \label{eq:commutpauli}
i [A, B] = i [\vec a \cdot \vec \sigma, \vec b \cdot \vec \sigma] = 2 ( \vec b \times \vec a) \cdot \vec \sigma,
\ee
where $\vec a$ and $\vec b$ denote the spinorial part of $A$ and $B$, respectively.

With the help of these notations, let us further concretize the physics of the system under investigation. In the following we denote operators with a hat, such as $\hat A$, and their integral kernels (matrix elements in the position space basis) without the hat, i.e. $A(x,y) = \matel{x}{\hat A}{y}$. The integral kernel of $\hat H_0$ is assumed to be of the general form $\matel{x}{ \hat H_0}{y}=H_0(x,y)\in\mH_2(\CC)$. Similarly, for the bath reference state $\hat \chi_B$ we write $\matel{Z}{\hat \chi_B}{Z'}=\chi_B(Z,Z') \in\bigotimes_{n=1}^N\mH_2(\CC)$, i.e. it is a hermitian $2^N\times 2^N$ matrix at every point $(Z,Z') \in \RR_Z^{Nd} \times \RR_Z^{Nd}$. The interaction Hamiltonian accounts for spin-dependent two particle interactions and is assumed to be diagonal in position space, $\matel{x,Z}{\hat H_I}{x',Z'} = H_I(x,Z,x',Z') = H_I(x,Z) \delta(x - x') \delta(Z - Z') \in\bigotimes_{n=1}^{N+1}\mH_2(\CC)$, where
\be \label{eq:interham}
H_I(x,Z) = \sum_{n = 1}^N \left ( \bigotimes_{k = 1}^{n-1} \un_k \right ) \otimes \mathbf{V}(x - z_{n} ) \otimes \left ( \bigotimes_{k = n+1}^N \un_k \right ).
\ee
Here, $\un_k$ denotes the $2 \times 2$ identity matrix referring to the $k$-th bath particle and $\mathbf{V}(r) \in \mH_2 \otimes \mH_2$ stands for the spin-dependent pair interaction potential that depends on the distance $r>0$ between the system's particle and one bath particle as well as on their spins. With the help of the multi-index notation, we can write the matrix elements of Eq. \eqref{eq:interham} as
\be
H_I^{\{\alpha \}\{ \beta \}}(x,Z) = \sum_{n = 1}^N \delta_{\alpha_1 \beta_1} \cdots \delta_{\alpha_{n-1} \beta_{n-1}} V_{\alpha_n \beta_n}(x - z_n) \delta_{\alpha_{n+1} \beta_{n+1}} \cdots \delta_{\alpha_N \beta_N},
\ee
where $\delta_{ij}$ denotes Kronecker's $\delta$ and we note that $V_{\alpha_n \beta_n}(r)$ are $2 \times 2$ matrices in the system's spin degrees of freedom which obey $V_{\alpha_n \beta_n}^\dagger(r) = V_{\beta_n \alpha_n}(r)$ due to the hermiticity of $\mathbf{V}(r)$. The particular form of the pair interaction $\mathbf{V}(r)$ is given by the system of interest and, therefore, depends on the type of particles or quasiparticles constituting the system and the environment. Independent of its actual form we may express $\mathbf{V}(r)$ with the help of Eqs. \eqref{eq:paulibase} as
\be \label{eq:vpauli0}
\mathbf{V}(r) = V_0(r) \otimes \un + \vec V(r) \odot \vec \sigma,
\ee
where $\vec V(r) \odot \vec \sigma = \sum_{i = 1}^3 V_i(r) \otimes \sigma_i$ and we defined the hermitian matrices
\begin{subequations}\label{eq:vpauli}
\bea 
V_0(r) & = & \frac{1}{2} \left [V_{11}(r) + V_{22}(r)\right ], \\
V_1(r) & = & \frac{1}{2} \left [V_{12}(r) + V_{12}^\dagger(r) \right], \\
V_2(r) & = & \frac{i}{2} \left [ V_{12}^\dagger(r) - V_{12}(r) \right ], \\
V_3(r) & = & \frac{1}{2} \left [ V_{11}(r) - V_{22}(r) \right ].
\eea
\end{subequations}
Here, we already employed that $V_{21}(r) = V_{12}^\dagger(r)$ and, hence, $V_i(r) \in \mH_2(\CC)$. We shall frequently employ the representation \eqref{eq:vpauli0} in what follows.

Hence, we regard the dynamics of a single quantum particle in contact with its environment. It is assumed that the dynamics are described in a proper fashion by the Lindblad Eq. \eqref{eq:hier} where the matrix element of the free particle Hammiltonian $\hat H_0$ is of the most general form $\matel{x}{\hat H_0}{y} = H_0(x,y)$ and the interaction $H_I$ between the system's particle and the environment, Eq. \eqref{eq:interham}, is composed of two particle interactions $\mathbf{V}(r)$ which is a function of the distance $r$ between the interacting particles.

\section{Integral Kernel of the dissipator} \label{sec:intkern}

It is the aim of this section to rewrite Eq. \eqref{eq:hier} as an equation for the integral kernel $\rho(x,y,t) = \matel{x}{\hat \rho(t)}{y} \in \mH_2(\CC)$ in order to prepare the transformation of Eq. \eqref{eq:hier} into the Wigner representation in Sec. \ref{sec:wigner}. Starting from Eq. \eqref{eq:hier}, a straight-forward calculation gives
\be \label{eq:hier1}
\partial_t \rho(x,y,t) + \frac{i}{\hbar} \cL \left ( H^{mf}_0 \right )(x,y,t) =  \cQ(\rho)(x,y,t),
\ee
where we defined the integral kernel of the commutator
\be \label{eq:comm}
\cL(A)(x,y,t) = \matel{x}{[\hat A, \hat \rho ]}{y} =  \int \dint{x'} \left [ A(x,x') \rho(x',y,t) - \rho(x,x',t) A(x',y) \right ],
\ee
for some general operator $A(x,x') = \matel{x}{\hat A}{x'} \in \mH_2(\CC)$ and the integral kernel of the dissipator
\be
\cQ(\rho)(x,y,t) = \matel{x}{\cD(\hat \rho)}{y}.
\ee
We note that Eq. \eqref{eq:hier1} is entirely equivalent to Eq. \eqref{eq:hier} and will serve as basis to derive the Wigner equation in Sec. \ref{sec:wigner}. The particular form of the dissipator $\cQ$ as well as of the matrix element of the mean-field Hamiltonian $H_{mf}(x)$ in Eq. \eqref{eq:hier1} will be determined in what follows. For this sake, we insert into Eq. \eqref{eq:dissi} the definition $\hat H^{mf}_I = \hat H_I - \hat H_{mf}$, see Sec. \ref{sec:notation}, in order to obtain
\bea \label{eq:dissimed}
\cD(\hat \rho) & = & \frac{\tau_0}{\hbar^2} \left [ \trB{2 \hat H_I \hat \rho \otimes \hat \chi_B \hat H_I - \hat H_I \hat H_I \hat \rho \otimes \hat \chi_B - \hat \rho \otimes \hat \chi_B \hat H_I \hat H_I} \right. \notag \\
&& \left.- 2 \hat H_{mf} \hat \rho \hat H_{mf} + \hat H_{mf} \hat H_{mf} \hat \rho + \hat \rho \hat H_{mf} \hat H_{mf} \right ].
\eea
To further simplify expression \eqref{eq:dissimed}, we note that inserting the interaction Hamiltonian \eqref{eq:interham} gives contributions in which we can take the partial trace over all but one or two bath particles. It is therefore advantegeous to define the partial traces of $\hat \chi_B$ over all but one or two bath particles, respectively, as
\begin{subequations}
\be \label{eq:densdef}
\chi^{(1)}_{ \beta_n \alpha_n } (z_n) := \sum_{ \{ \alpha \}_n  } \sinto{Z_n} \chi_B^{ \{ \alpha,\beta_n \} \{ \alpha \} }(Z,Z),
\ee
\be \label{eq:chi2}
 \chi^{(2)}_{ \beta_n\gamma_m\alpha_n\beta_m }(z_n,z_m) :=  \sum_{ \{ \beta \}_{nm}  } \sinto{Z_{nm}} \chi_B^{ \{ \beta,\gamma_m \} \{ \beta,\alpha_n \} }(Z,Z)\,,
\ee
where the indices $n,m$ still refer to specific particles. Since we assume that the bath particles are indistinguishable (Sec. \ref{sec:notation}), we may omit these indices. It follows immediately that the corresponding matrices $\chi^{(1)}(z) \in \mH_2(\CC)$ and $\chi^{(2)}(z,z') \in \mH_2(\CC) \otimes \mH_2(\CC)$ are normalized, i.e.
\be
\sum_{\alpha} \int \dint{z} \chi^{(1)}_{\alpha \alpha}(z) = 1, \qquad \sum_{\alpha \beta} \int \dint{z'} \dint{z'} \chi^{(2)}_{\alpha \beta \alpha \beta} (z,z')= 1.
\ee
\end{subequations}
Moreover, due to the indistinguishability it turns out to be beneficial to define the one- and two-particle spin-density matrices $\mathrm{N}^{(1)}(z) \in \mH_2( \CC)$ and $\mathbf{N}^{(2)}(z,z') \in \mH_2(\CC) \otimes \mH_2(\CC)$ with matrix elements
\begin{subequations}\label{replace1}
\bea
n^{(1)}_{\alpha \beta}(z) & := & N \chi^{(1)}_{\alpha \beta}(z), \\
n^{(2)}_{\alpha \alpha' \beta \beta'}(z,z') & : = & N^2\chi^{(2)}_{\alpha \alpha' \beta \beta'}(z,z'),
\eea
\end{subequations}
respectively. Let us briefly comment on this definitions: Employing the notation \eqref{eq:paulibase}, we call
\be \label{eq:npauli}
\mathrm{N}^{(1)}(z) = n_0(z) \un + \vec n(z) \cdot \vec \sigma,
\ee
the spinorial density of bath particles. Here the scalar part $n_0(z)$ is the density of bath particles at $z$ and the spin part $\vec n(z)$ is the directional spin polarization of the bath at $z$. The spin part $\vec n(z)$ is proportional to the magnetization $\vec m(z)$ of the environment\cite{white}, thus, the spin-density matrix $\mathrm{N}^{(1)}(z)$ at position $z$ contains the complete spin resolved information about the probability of finding a bath particle at position $z$. In a similar fashion we regard the matrix $\mathbf{N}^{(2)}(z,z')$ as the spin-resolved two-particle density matrix of the environment.

With the help of the above definitions a straight forward calculation allows to express the mean-field interaction Hamiltonian $H_{mf}(x)$,
\be \label{eq:Hmf3}
H_{mf}(x) = \sinto{z} V_{\alpha \beta}(x-z) n^{(1)}_{ \beta\alpha }(z)\,,
\ee
as well as the integral kernel of the dissipator (see App. \ref{app:dissi1})
\be \label{eq:dissi2}
\begin{aligned}
\cQ(\rho)(x,y,t) =& \frac{2 \tau_0}{\hbar^2} \dinto{z}{z'} \kappa_{\beta \beta' \alpha \alpha'}(z,z') \Big[ V_{\alpha \beta}(x-z)\rho(x,y,t)V_{\alpha'\beta'}(y-z')\Big. \\
&\Big.- \frac{1}{2}V_{\alpha \beta}(x-z)V_{\alpha'\beta'}(x-z')\rho(x,y,t) - \frac{1}{2}\rho(x,y,t)V_{\alpha \beta}(y-z)V_{\alpha'\beta'}(y-z') \Big],
\end{aligned}
\ee
for the system under investigation. Here and in what follows we shall employ Einstein's sum convention. Moreover, $\kappa_{\alpha \alpha' \beta \beta'}(z,z')$ are the matrix elements of the modified density-density covariance matrix $\mathbf{K}(z,z')$ defined via
\be \label{eq:kappadef}
\mathbf{K}(z,z') = \mathbf{C}(z,z') + \mathbf{D}(z,z'),
\ee
where $\mathbf{D}(z,z') \in \mH_2(\CC) \otimes \mH_2(\CC)$ is given via $d_{\alpha \alpha' \beta \beta'}(z,z') := n^{(1)}_{\alpha \beta}(z) \delta_{\alpha \alpha'} \delta_{\beta \beta'} \delta(z - z')$ and we introduced the spinorial density-density covariance matrix (or environmental covariance matrix)
\be \label{eq:defc}
\mathbf{C}(z,z') = \mathbf{N}^{(2)}(z,z') - \mathrm{N}^{(1)}(z) \otimes \mathrm{N}^{(1)}(z').
\ee

Hence, we determined all components of the evolution equation of the integral kernel $\rho(x,y,t)$, Eq. \eqref{eq:hier1}. The particular form of the integral kernel of the commutator as well as of the dissipator were determined for the interaction potential \eqref{eq:interham} to be of the forms \eqref{eq:comm} and \eqref{eq:dissi2}. The action of the dissipator $\cQ$ on $\rho$ is, thus, determined by the matrix elements of the modified spin-resolved density-density covariance matrix $\mathbf{K}(z,z')$ \eqref{eq:kappadef} and the interaction potential $\mathbf{V}(r)$. Furthermore, the explicit form of the mean-field Hamiltonian $H_{mf}(x)$ was obtained, Eq. \eqref{eq:Hmf3}. The derivation of the Wigner equation in Sec. \ref{sec:wigner} will essentially be based on these equations.

However, before proceeding to the next section let us briefly discuss the integral kernel of the mean-field interaction, $H_{mf}(x)$, Eq. \eqref{eq:Hmf3}, in more detail. The mean-field interaction $H_{mf}(x)$ can be regarded as the partial trace $\mathrm{tr}_1 ( \cdot )$ over the bath's particle spin degrees of freedom of the convolution of the potential $\mathbf{V}(r)$, Eq. \eqref{eq:interham}, with the spinorial density $\mathrm{N}^{(1)}(z)$,
\bea \label{eq:hmf4}
H_{mf}(x) & = & \int \dint{z} \mathrm{tr}_1 \left [ \mathbf{V}( x - z ) \mathrm{N}^{(1)}(z)\right] \notag \\
 & = & \int \dint{z} V_i(x - z) n_i(z),
\eea
where the sum runs from $i = 0,\ldots,3$ and the matrices $V_i(r) \in \mH_2(\CC)$ have been defined in Eq. \eqref{eq:vpauli}.

The mean field interaction \eqref{eq:hmf4} is easily concretized for two particularly interesting physical situations. As a first example we regard the case that the interaction is spin independent, i.e.
\be \label{eq:vcasea}
\mathbf{V}( r  ) = v( r  ) \mathds{1} \otimes \mathds{1},
\ee
where $v(r) \in \RR$ and we obtain the familiar expression
\be \label{eq:casea}
H_{mf}(x) = \int \dint{z} v( x - z)  n_0(z) \mathds{1},
\ee
which is known from spin independent mean-field theory\cite{ashcroft}. Here the mean-field interaction is substantially determined by the total density of bath particles $n_0(z)$ at position $z$ and, therefore, insensible for any spin-polarization of the environment. On the other hand, in the case that
\be \label{eq:vcaseb}
\mathbf{V}(  r ) = v ( r  ) \sigma_3 \otimes \sigma_3,
\ee
i.e. a distance-dependent $3$-polarized spin-spin interaction, we have
\be\label{eq:caseb}
H_{mf}(x) = \int \dint{z} v( x - z)  n_3(z) \sigma_3,
\ee
which is a mean-field interaction that depends on the $3$-spin polarization of the environment $n_3(z)$ at position $z$. Since the spin polarization of the bath may be connected to a magnetization $m_3(z)$\cite{white} via $m_3(z) \propto n_3(z)$, we interpret Eq. \eqref{eq:caseb} as the interaction of the system's spin with a position dependent effective magnetic field in $3$-direction. Since we may write the mean-field interaction as $H_{mf}(x) = h^{mf}_3(x) \sigma_3$ we regard $h_3^{mf}(x)$ as the $3$-component of the effective magnetic field at position $x$. Here, the $3$-component of the effective magnetic field $h^{mf}_3(x)$ is given by a convolution between $v(x - z)$ and the magnetization $m(z) \propto n_3(z)$.

We shall come back to these two particular examples in the course of the following sections.


\section{Wigner representation of the Lindblad equation} \label{sec:wigner}

It is the aim of this section to apply the Wigner transform $\cW$ to Eq. \eqref{eq:hier1}. The ensuing transport equation for $W = \cW(\rho)$ is a spinorial Wigner equation equipped with a quantum collision operator stemming from the dissipator $\cQ$ defined in Eq. \eqref{eq:dissi2}. This result will serve as a starting point for the semiclassical analysis performed in Sec. \ref{sec:scalenlimit}.

In what follows we shall refrain from an explicit notation of the time argument $t$. Let us define the Wigner transform of $\rho(x,y) \in \mH_2(\CC)$ and its inverse. The element-wise Wigner transform reads
\be \label{def:wignerphys}
w_{ij}(x,\eta) : = \cW(\rho_{ij}) =\frac{1}{(2\pi\hbar)^d}\sinto{y}{\rho_{ij}\left(x+\frac{y}{2},x-\frac{y}{2}\right) \exp \left (-\frac{i}{\hbar} y \cdot \eta \right )}\,,
\ee
and the corresponding inverse transform is given by
\be \label{invwig}
\rho_{ij}(x,y) = \cW^{-1}(w_{ij}) = \sinto{\eta}{w_{ij}\left(\frac{x+y}{2}, \eta\right)\exp \left [\frac{i}{\hbar} (x - y) \cdot \eta \right ]}.
\ee
By convention, we shall understand all Wigner transforms in an element-wise fashion as defined in Eq. \eqref{def:wignerphys} and \eqref{invwig}, i.e. we denote $W = \cW(\rho)$. We remark that the Wigner function $W$ must not be interpreted as a phase space distribution function since its eigenvalues also take on negative values. The size of negative regions of the Wigner function display the wavefunction's ability to interfere and may be used as a measure for the non-classicality of the system.\cite{kenfack}

We now apply the Wigner transform \eqref{def:wignerphys} to Eq. \eqref{eq:hier1} for the integral kernel $\rho$. For this sake, let $\mathfrak{a}(x,\eta) = (2 \pi \hbar)^d \cW[A(x,y)] \in \mH_2(\CC)$ denote the phase space symbol of an operator $\hat A$ acting in the system's Hilbert space $\cH$. Then we write for any phase space symbol $\mathfrak{a}$ the Moyal bracket \cite{zachos} as (see App. \ref{app:moyal})
\bea \label{eq:commterm}
\cL_\hbar \left ( \mathfrak{a} \right ) (x,\eta,t) & : = & \cW \left [ \cL (A) \right ] (x, \eta,t) \notag \\
& = & \frac{1}{(2 \pi\hbar)^{2d}} \int \dint{z} \dint{z'} \int \dint{\xi} \dint{\xi'} \left [ \mathfrak{a} \left ( z' + \frac{1}{4} z, \xi + \frac{1}{4} \xi' \right ) \right. \notag \\
&& \left.  W \left ( z' - \frac{1}{4}z, \xi - \frac{1}{4} \xi',t\right ) - W \left ( z' + \frac{1}{4} z, \xi + \frac{1}{4} \xi',t \right ) \right. \notag \\
&& \left.  \mathfrak{a} \left ( z' - \frac{1}{4}z, \xi - \frac{1}{4} \xi' \right ) \right ] \exp \left [  \frac{i}{\hbar} \xi' \cdot (x - z') \right ] \notag \\
&& \times \exp \left [ - \frac{i}{\hbar} z \cdot \left ( \eta - \xi \right ) \right ],
\eea
which is the Wigner transform of the integral kernel of the commutator, $\cL(A)$ defined in Eq. \eqref{eq:comm}. In the general case, we remember that $H^{mf}_0(x,y) = H_0(x,y) + H_{mf}(x,y)$, where the mean-field interaction $H_{mf}(x,y) = H_{mf}(x) \delta(x-y)$ was concretized in Eq. \eqref{eq:Hmf3}. Let $\mathfrak{h}_0(x,\eta) = (2 \pi \hbar)^d \cW( H_0)$ denote the symbol of $H_0$ in phase space and let $\mathfrak{h}^{mf}_0 = \mathfrak{h}_0 + H_{mf}(x)$ be the phase space symbol of $H^{mf}_0$. Then the Wigner-transformed Eq. \eqref{eq:hier1} may be written in a compact form as
\be \label{eq:wignerhier1}
\partial_t W + \frac{i}{\hbar} \cL_\hbar ( \mathfrak{h}^{mf}_0) = \frac{\tau_0}{\hbar^2} \cQ_\hbar(W).
\ee
where the Wigner representation of the dissipator \eqref{eq:dissi2} reads
\bea \label{eq:dissiwig}
\cQ_\hbar(W) &  : = & \cW \left [ \cQ (\rho) \right ]  \notag \\
 & = & \frac{2}{(2 \pi \hbar)^d}\int \dint{x'} \dint{z} \dint{z'} \int \dint{\eta'} \kappa_{\beta \beta' \alpha \alpha'}(z, z') \notag \\
&& \times \left [ V_{\alpha \beta} \left ( x + \frac{1}{2} x' - z \right ) W' V_{\alpha'\beta' } \left ( x - \frac{1}{2} x' - z' \right )  \right. \notag \\
&& - \frac{1}{2} V_{\alpha\beta } \left ( x + \frac{1}{2} x' - z \right ) V_{\alpha'\beta' } \left ( x + \frac{1}{2} x' - z' \right ) W' \notag \\
&& \left. -\frac{1}{2} W' V_{\alpha\beta } \left ( x - \frac{1}{2} x' - z \right )V_{\alpha'\beta' } \left ( x - \frac{1}{2} x' - z' \right ) \right ] \notag \\
&& \times \exp \left [- \frac{i}{\hbar} x' \cdot ( \eta - \eta') \right ],
\eea
with $W' = W(x,\eta',t)$.

Eq. \eqref{eq:wignerhier1} is the Wigner equation for a single spin-$1/2$ particle in interaction with its environment and is entirely equivalent to Eq. \eqref{eq:hier1} of operator symbols in position space, from which it was derived. The lhs of Eq. \eqref{eq:wignerhier1} represents the free flight of this particle and is therefore analogous to the equation obtained from the von Neumann equation in the Wigner picture, except for the mean-field correction. For instance, if $H^{mf}_0(x,y)$ gives the symbol $\mathfrak{h}^{mf}_0(x,\eta) = \left [ \eta^2 / (2m) + u(x) \right ] \un + \vec \Omega(x,\eta) \cdot \vec \sigma + H_{mf}(x)$, with the particle's mass $m$, the scalar external potential $u(x)$ and the exchange field $\vec \Omega(x,\eta) \cdot \vec \sigma$, one obtains the free flight term of the Vlasov equation
\be \label{eq:vlasov}
\frac{i}{\hbar} \cL_\hbar \left ( \mathfrak{ h}^{mf}_0 \right ) = \frac{\eta}{m} \cdot \nabla_x W+ \frac{i}{\hbar} \cL_\hbar \left (u \un \right ) + \frac{i}{\hbar} \cL_\hbar \left ( \vec \Omega\cdot \vec \sigma + H_{mf} \right ).
\ee
Here, $\cL_\hbar \left ( u \un \right )$ is of the well known form (see App. \ref{app:moyal})
\bea \label{eq:vlasov2}
\cL_\hbar(u \un) & = & \frac{1}{(2 \pi \hbar)^d} \int \dint{x'} \int \dint{\eta'} \left [ u \left ( x + \frac{x'}{2} \right ) - u \left ( x - \frac{x'}{2} \right ) \right ] W' \notag \\
 && \times \exp \left [ - \frac{i}{\hbar} x' \cdot (\eta - \eta') \right ].
\eea
The mean-field correction in Eq. \eqref{eq:vlasov} is a first contribution stemming from the interaction between the system's particle and the environment and its particular action depends on the form of the two particle interaction $\mathbf{V}(r)$. As illustrated in Eqs. \eqref{eq:casea} and \eqref{eq:caseb} the mean-field interaction may give rise to a scalar as well as to a spinorial contribution to the one-particle Hamiltonian $H_0(x,y)$. In particular, if $\mathbf{V}(r)$ is spin-independent, i.e. of the form Eq. \eqref{eq:vcasea}, the mean field contribution to Eq. \eqref{eq:vlasov} is of the form Eq. \eqref{eq:vlasov2} and $H_{mf}(x)$ behaves like an external scalar potential $u(x)$. On the other hand, if $\mathbf{V}(r)$ models a spin-dependent interaction of the form Eq. \eqref{eq:vcaseb} a brief calculation demonstrates that the mean field contribution to Eq. \eqref{eq:vlasov} takes on the form
\bea
\cL_\hbar \left (h^{mf}_3 \sigma_3 \right ) & = & \int \dint{\eta'} \tilde h^{mf}_3(\eta') \left [\sigma_3 W \left (x, \eta - \frac{\eta'}{2} ,t \right ) - W \left ( x, \eta + \frac{\eta'}{2}, t \right ) \sigma_3 \right ] \notag \\
 && \times \exp \left ( \frac{i}{\hbar} x \cdot \eta' \right ).
\eea
Here, $\tilde h^{mf}_3(\eta')$ denotes the Fourier transform of the $3$-component of the vector $\vec h_{mf}(z)$, see App. \ref{app:fourier}. This term may be interpreted in the sense that it is accounting for the mixing of momentum components of different spin species due to the effective magnetic field $h^{mf}_3(z)$ induced by the spin degrees of freedom of the environment (Sec. \ref{sec:intkern}).

The rhs of Eq. \eqref{eq:wignerhier1} is a quantum mechanical collision integral (or collision operator) for the Wigner equation. We note that Eq. \eqref{eq:wignerhier1} together with Eqs. \eqref{eq:commterm} and \eqref{eq:dissiwig} will serve as a starting point for the semiclassical analysis carried out in Sec. \ref{sec:scalenlimit}, however, let us comment on some general properties of the quantum collision operator first.

\section{The quantum collision operator} \label{sec:coll}

We discuss some general features of the collision operator $\cQ_\hbar(W)$ acting on the spinorial Wigner function $W$ according to Eq. \eqref{eq:dissiwig} in order to understand its action in more detail. This will prove to be crucial in the semiclassical analysis of Eqs. \eqref{eq:wignerhier1}, \eqref{eq:commterm} and \eqref{eq:dissiwig} in Sec. \ref{sec:scalenlimit}. We insert inverse Fourier transforms (see App. \ref{app:fourier}) of the interaction potential,
\begin{subequations}
\be \label{eq:fourierV}
V_{\alpha \beta}(x) = \int \dint{\xi} \tilde V_{\alpha \beta}(\xi) \exp \left ( \frac{i}{\hbar} x \cdot \xi \right ),
\ee
and of the modified spin resolved density-density covariance,
\be
\kappa_{\beta \beta' \alpha \alpha'}(z, z') = \int \dint{\xi} \dint{\xi'} \tilde \kappa_{\beta \beta' \alpha \alpha'}(\xi, \xi') \exp \left [ \frac{i}{\hbar} \left ( z \cdot \xi + z' \cdot \xi' \right ) \right ],
\ee
\end{subequations}
into Eq. \eqref{eq:dissiwig} and subsequently integrate over $x',z$ and $z'$ in order to obtain
\bea \label{eq:quantcoll}
\cQ_\hbar (W)(x, \eta,t) & = & 2 (2 \pi \hbar)^{2d} \int \dint{\xi} \dint{\xi'} \tilde \kappa_{\beta \beta' \alpha \alpha'}(\xi, \xi') \notag \\
&& \times \left [ \tilde V_{\alpha\beta } \left ( \xi \right ) W \left (x, \eta - \frac{\xi - \xi'}{2} ,t \right ) \tilde V_{\alpha'\beta' } \left (\xi' \right )  \right. \notag \\
&& - \frac{1}{2} \tilde V_{\alpha\beta } \left ( \xi \right ) \tilde V_{\alpha'\beta' } \left ( \xi' \right ) W \left (x, \eta - \frac{\xi + \xi'}{2},t \right ) \notag \\
&& \left. -\frac{1}{2} W \left (x, \eta + \frac{\xi + \xi'}{2},t \right ) \tilde V_{\alpha\beta } \left ( \xi \right ) \tilde V_{\alpha'\beta' } \left ( \xi' \right ) \right ] \notag \\
&& \times \exp \left [\frac{i}{\hbar} x \cdot (\xi + \xi' ) \right ].
\eea
We denote this as the representation of the collision operator \eqref{eq:dissiwig} as a momentum space integral. Whether or not this representation is more convenient than Eq. \eqref{eq:dissiwig} depends on the system under investigation. However, we shall now discuss the special case that $\mathbf{K}(z,z') = \mathbf{K} (z - z')$ and it will turn out that in this case representation \eqref{eq:quantcoll} is the more convenient one. We stress that it has to be checked carefully whether or not the simplification $\mathbf{K}(z ,z') = \mathbf{K}(z - z')$ is valid for the system under investigation. For the sake of a simplified discussion, we shall assume it to be a valid approximation in what follows.

From definition \eqref{eq:kappadef} we remember that $\mathbf{K}(z,z') = \mathbf{C}(z,z') + \mathbf{D}(z,z')$. For translational invariant systems the covariance $\mathbf{C}(z,z') = \mathbf{C}(z - z')$. If we further restrict our discussion to the case of a constant particle distribution, i.e. $\mathrm{N}^{(1)}(z) = \mathrm{const}$, we also have $\mathbf{D}(z,z') = \mathbf{D}(z - z')$. It is, therefore, a sufficient condition to assume a space homogeneous environment whose spinorial density is constant in space in order to justify the above simplification. Hence, we regard the case that $\tilde{\mathbf{K}}(\xi,\xi') = \tilde{\mathbf{K}} \left ( \frac{\xi + \xi'}{2} \right ) \delta(\xi + \xi')$, see App. \ref{app:fourier}, and, therefore, Eq. \eqref{eq:quantcoll} simplifies to
\bea \label{eq:dissiwig2}
\cQ_\hbar (W)(x, \eta,t) & = & 2 (2 \pi \hbar)^{2d} \int \dint{\eta'} \tilde \kappa_{\beta \beta' \alpha \alpha'}(\eta - \eta') \notag \\
&& \times \left [ \tilde V_{\alpha\beta } \left ( \eta - \eta' \right ) W' \tilde V_{\alpha'\beta' } \left (\eta - \eta' \right )  \right. \notag \\
&& - \frac{1}{2} \tilde V_{\alpha\beta } \left (\eta - \eta' \right ) \tilde V_{ \alpha'\beta'} \left (\eta - \eta' \right ) W \notag \\
&& \left. -\frac{1}{2} W  \tilde V_{\alpha\beta } \left ( \eta -\eta' \right )\tilde V_{\alpha'\beta' } \left ( \eta - \eta' \right ) \right ],
\eea
with $W = W(x,\eta,t)$ and $W' = W(x,\eta',t)$. Please note that we employed that $\tilde V_{\alpha \beta}(\xi) = \tilde V_{\alpha \beta}(-\xi)$ according to Eq. \eqref{eq:interham}. In what follows we shall highlight some general features of the quantum collision operator \eqref{eq:dissiwig2}.

It is demonstrated in App. \ref{app:rewrite} that we may rewrite Eq. \eqref{eq:dissiwig2} in the even more convenient form
\bea \label{eq:dissiwig3}
\cQ_\hbar (W)(x, \eta,t) & = & 2 (2 \pi \hbar)^{2d} \int \dint{\eta'}\rho_{i}(\eta - \eta') \notag \\
&& \times \left [ S_{i} \left (\eta - \eta' \right ) W' S_{i} \left (\eta-\eta' \right )  \right. \notag \\
&& - \frac{1}{2} S_{i} \left ( \eta -\eta' \right )  S_{i} \left (\eta - \eta' \right ) W \notag \\
&& \left. -\frac{1}{2} W  S_{i} \left ( \eta -\eta' \right ) S_{i} \left ( \eta - \eta' \right ) \right ],
\eea
i.e. as a sum of collision operators of the form of the Boltzmann equation, Eq. \eqref{QF}. Hence, in the particular case that the modified covariance fulfills $\mathbf{K}(z,z') = \mathbf{K}(z - z')$, the quantum collision operator \eqref{eq:dissiwig} turns out to be of the form proposed for the semiclassical SBE, Eq. \eqref{QF}. However, since $W$ can take negative eigenvalues it is not a collision integral in the classical sense as already emphasized at the beginning of this section. Here, $\rho_i(\eta') \in \RR$, $S_i(\eta') \in \mH_2(\CC)$ and $i = 0,\ldots,3$. As demonstrated in App. \ref{app:rewrite} the hermitian matrices $S_i$ are linear combinations of the matrices $V_i$, where the weights are determined by the matrix $\tilde{\mathbf{K}}(\eta')$. The scalar functions $\rho_i(\eta') \in \RR$ are the eigenvalues of $\tilde{\mathbf{K}}(\eta')$.

Let us briefly clarify the terminology used in the subsequent analysis: According to Eq. \eqref{eq:paulibase} we may express $W$ as $W = w \un + \vec w \cdot \vec \sigma$, where we denote by $w = 1 / 2~ \tr{W}$ the scalar part and by $\vec w = 1/2~ \tr{W \vec \sigma}$ the spin part of $W$. Furthermore, we call $N = \int \dint{\eta} W$ the spinorial particle density (or distribution) and $M = \int \dint{x} W$ the spinorial momentum distribution, which may also be decomposed according to Eq. \eqref{eq:paulibase}. Again, the corresponding terms are referred to as the scalar and the spin part of $N$ and $M$, respectively.

The collision operator \eqref{eq:dissiwig3} may be decomposed according to
\be
\cQ_\hbar ( \cdot) = \cQ_\hbar^{(1)} ( \cdot) + \cQ_\hbar^{(2)} ( \cdot ),
\ee
where
\bea \label{eq:dissiwigc1}
\cQ_\hbar^{(1)} ( W ) & = &  2(2 \pi \hbar)^{2d} \int \dint{\eta'} \rho_i(\eta -\eta') S_{i} \left ( \eta - \eta' \right ) \left [ W'  - W  \right ] S_{i} \left (\eta -\eta' \right ).
\eea
and
\bea \label{eq:dissiwigc2}
\cQ_\hbar^{(2)} (W)(x, \eta,t) & = & (2 \pi \hbar)^{2d} \int \dint{\eta'} \rho_i(\eta')  \left [ \left [ S_{i} \left ( \eta' \right ), W \right ], S_{i} \left (\eta' \right )  \right ].
\eea
We note that $\cQ_\hbar^{(1)}$ is of the general master equation form '\textit{gain term $ - $ loss term}'. It is easily observed that we have
\be \label{eq:momrelax}
\int \dint{\eta} \cQ_\hbar^{(1)}(W) = 0,
\ee
for arbitrary $W$, i.e. the spinorial particle density $N$ is conserved by $\cQ_\hbar^{(1)}$. To be more specific, the particle distribution $N(x,t)$ is not affected by the action of $\cQ_\hbar^{(1)}$ while the scalar as well as the spin part of the momentum distribution $M(\eta,t)$ are changed.

In a similar fashion we obtain for $\cQ_\hbar^{(2)}$ that
\be \label{eq:spindec}
\mathrm{tr} \left [\cQ_\hbar^{(2)}(W) \right ] = 0,
\ee
for arbitrary $W$, i.e. the scalar part $w$ of the Wigner function $W$ is conserved by $\cQ_\hbar^{(2)}$. Thus $\cQ_\hbar^{(2)}$ acts solely on the spin part $\vec w$. Interestingly, the conservation of $w$ signifies that both, the scalar particle distribution $n(x,t)$ as well as the scalar momentum distribution $m(\eta,t)$ remain unaffacted. The operator $\cQ_\hbar^{(2)}$ may therefore be identified as accounting solely for local spin-flip processes.

Combining Eq. \eqref{eq:spindec} and Eq. \eqref{eq:momrelax} gives the property
\be
\int \dint{\eta} \mathrm{tr} \left [ \cQ_\hbar (W) \right ] = 0,
\ee
for arbitrary $W$. Thus the scalar part of the spinorial particle density, $n(x,t)$ is conserved by $\cQ_\hbar$ as in the well-known scalar case.\cite{cercignani} However, the spin part and in particular the local spin polarization $\vert \vec n(x,t) \vert$ is changed, i.e. $\cQ_\hbar$ accounts for spin-flip processes due to its part $\cQ_\hbar^{(2)}$.

We now investigate in more detail the operator $\cQ_\hbar^{(2)}$. Due to its particular form \eqref{eq:dissiwigc2} we may regard the arguments $x,\eta$ and $t$ of $W$ as parameters and assume that the domain of $\cQ_\hbar^{(2)}$ is $\mH_2(\CC)$ in the following analysis. At first, we are interested in its kernel denoted by $\mathrm{Ker}(\cQ_\hbar^{(2)})$,  because, as will become apparant in what follows, we may under certain premises reason from the structure of $\mathrm{Ker}(\cQ_\hbar^{(2)})$ on the stationary spin distribution of $W$ for some given $\mathbf{V}(r)$. The $\mathrm{Ker}(\cQ_\hbar^{(2)})$ is defined as the set of all matrices $A \in \mH_2(\CC)$ for which $\cQ_\hbar^{(2)}(A) = 0$. As demonstrated in App. \ref{app:dissicomm} we have
\be \label{eq:q2cond}
\cQ_\hbar^{(2)} (A) = 0 \Leftrightarrow [S_{i}(\eta'),A] = 0, \qquad \forall i \quad \text{ and } \quad \forall \eta',
\ee
if detailed balance is requred for $A$. Since $\cQ_\hbar^{(2)}$ is linear and self-adjoint with respect to the Hilbert-Schmitdt scalar product $\tr{AB}$, $A,B \in \mH_2(\CC)$, its domain $D(\cQ_\hbar^{(2)})$ may be decomposed into
\be \label{eq:domain}
D(\cQ_\hbar^{(2)}) = \mH_2(\CC) = \mathrm{Ker}(\cQ_\hbar^{(2)}) \oplus \mathrm{Ker}(\cQ_\hbar^{(2)})^\bot,
\ee
where
\be
\mathrm{Ker}(\cQ_\hbar^{(2)})^\bot : = \left \{ A \in \mH_2(\CC) \vert \tr{AB} = 0 \forall B \in \mathrm{Ker}(\cQ_\hbar^{(2)}) \right \},
\ee
is the space orthogonal to $\mathrm{Ker}(\cQ_\hbar^{(2)})$.

If we knew the projection of some particular state $W$ on $\mathrm{Ker}(\cQ_\hbar^{(2)})$, we could assume that $\cQ_\hbar^{(2)}$ can be approximated by a relaxation time ansatz
\be \label{eq:relaxapprox}
\overline \cQ_\hbar^{(2)}(W) := \frac{\overline W - W}{\tau_2},
\ee
where $\overline W$ denotes the projection of $W$ on the kernel of $\cQ_\hbar^{(2)}$ for some given $\mathbf{V}(r)$ and $\tau_2$ is the mean relaxation time. The action of the operator $\overline \cQ_\hbar^{(2)}$ as defined in Eq. \eqref{eq:relaxapprox} on some function $W$ is that it relaxes $W$ into $\overline W$, where $\tr{W} = \tr{\overline W}$ since $\tr{\cQ_\hbar^{(2)}} = 0$. We stress that, again, whether or not Eq. \eqref{eq:relaxapprox} is a valid approximation has to be checked carefully for the problem of interest.

It is the aim of the following paragraphs to determine the equilibrium distribution $\overline W \in \mathrm{Ker} \left (\cQ^{(2)}_\hbar \right )$ for a given two particle interaction potential $\mathbf{V}(r)$. First of all, we deduce from Eq. \eqref{eq:q2cond} and \eqref{eq:commutpauli} that for $A \in \mathrm{Ker}(\cQ_\hbar^{(2)})$ we have
\be \label{eq:q2cond2}
[S_i(\eta'),A] = 2 i \left [ \vec s_i(\eta') \times \vec a \right ] \cdot \vec \sigma \stackrel{!}{=} 0.
\ee
Since $A$ does not depend on $\eta'$, this is for $\vec s_i(\eta') \neq 0$ and $\vec a \neq 0$ only possible if the direction of the spin part of $S_i(\eta')$ does not depenend on $\eta'$, i.e. $\vec s_i(\eta')/\vert \vec s_i(\eta') \vert = \mathrm{const}$. On the other hand, if it {\em does} depend on $\eta'$, i.e. if $\vec s_i(\eta')/\vert \vec s_i(\eta') \vert \neq \mathrm{const}$, the only possible solution to Eq. \eqref{eq:q2cond2} is $\vec a = 0$. In particular, it follows that in this case the spin part of $\overline W$ is zero. Hence, the unique projection $W \mapsto \overline W$ reads
\be \label{eq:spindepol}
\overline W = \frac{1}{2} \tr{W} \un,
\ee
i.e. the operator $\overline \cQ_\hbar^{(2)}$ destroys spin polarization on a finite time scale $\tau_2$ whenever the interaction potential's direction of the spin polarization $\vec s_i(\eta') / \vert \vec s_i(\eta') \vert \neq \mathrm{const}$. In a similar fashion, we note that Eq. \eqref{eq:spindepol} is the sole solution to Eq. \eqref{eq:q2cond} if the directions of the spin parts of $S_i$ are not identical for different $i$. We highlight that this mechanism of spin depolarization is not based on spin orbit interaction\cite{fabian1,fabian2} but on direct spin-spin interaction between different particles. Again, we emphasize: Whenever the direction of the spin part of the interaction $S_i$ is a function of $\eta'$ or not equivalent for different $i$, the action of $\overline \cQ_\hbar^{(2)}$ unavoidably leads to spin depolarization on a finite time scale, as described in Eq. \eqref{eq:spindepol}.

We now concentrate on the special case that the direction of the spin part of $S_i(\eta')$ is constant in $\eta'$ and identical for all $i$. If $\vec s_i(\eta') = 0$ then $\cQ_\hbar^{(2)} ( \cdot ) = 0$ and the operator $\cQ_\hbar$ does not account for spin-decoherence at all. However, if $\vec s_i(\eta') \neq 0$ then $\vec s_i(\eta') / \vert \vec s_i(\eta') \vert = \mathrm{const}$ is identical for all $i$, i.e. we may write
\be
\vec s_i(\eta') = \gamma_{i}(\eta') \vec \lambda,
\ee
where $\gamma_{i}(\eta')$ is some scalar function and $\vert \vec \lambda \vert = 1$. Hence,
\be
S_i(\eta') = s_i(\eta') \un + \gamma_{i}(\eta') \vec \lambda \cdot \vec \sigma, \qquad \vert \vec \lambda \vert = 1.
\ee
Let $\Sigma$ be the matrix which diagonalizes $\vec \lambda \cdot \vec \sigma$ and, according to Eq. \eqref{eq:q2cond}, also $\overline W \in \mathrm{Ker}(\cQ_\hbar^{(2)})$. We then obtain that the unique equilibrium spin configuration $\overline W$ reads
\be \label{eq:kernelc2}
\overline W = \frac{1}{2} \tr{W} \un + \frac{1}{2} \tr{\sigma_3 \Sigma^\dagger W \Sigma} \Sigma \sigma_3 \Sigma^\dagger.
\ee

In order to prove the result Eq. \eqref{eq:kernelc2} we note that we may decompose an arbitrary matrix $W$ according to $W = \mD(W) + \mO(W)$, where $\mD(W)$ denotes the matrix whose off-diagonal elements are zero and $\mO(W)$ denotes the matrix whose diagonal elements are zero. Now, according to Eq. \eqref{eq:q2cond} we have
\be
\overline W \in \mathrm{Ker}(\cQ_\hbar^{(2)}) \Leftrightarrow \Sigma^\dagger \overline W \Sigma = \mD(\Sigma^\dagger \overline W \Sigma),
\ee
is solely diagonal. In a similar fashion
\be
W' \in \mathrm{Ker}(\cQ_\hbar^{(2)})^\bot \Leftrightarrow \Sigma^\dagger W' \Sigma = \mO(\Sigma^\dagger W' \Sigma),
\ee
is solely off-diagonal.\footnote{This statement is is easily proved: suppose $\mD(\Sigma^\dagger W' \Sigma) \neq 0$. We may then define a matrix $A = \Sigma \mD(\Sigma^\dagger W' \Sigma) \Sigma^\dagger \in \mathrm{Ker}(\cQ_\hbar^{(2)})$, which is a contradiction to the assumption $W' \in \mathrm{Ker}(\cQ_\hbar^{(2)})^\bot$, which proofs '$\Leftarrow$'. On the other hand, let $W' \in \mathrm{Ker}(\cQ_\hbar^{(2)})^\bot$ and $B \in \mathrm{Ker}(\cQ_\hbar^{(2)})$. Then we have
\be
0 = \tr{W' B} = \tr{\Sigma^\dagger W'\Sigma \Sigma^\dagger B \Sigma} = \tr{\Sigma^\dagger B \Sigma \mD(\Sigma^\dagger B \Sigma) } \Rightarrow \Sigma^\dagger W' \Sigma = \mO(\Sigma^\dagger W' \Sigma), 
\ee
which demonstrates '$\Rightarrow$' and the proof is completed.} Hence, with the help of Eq. \eqref{eq:domain}, since $\Sigma^\dagger W \Sigma = \mD(\Sigma^\dagger W \Sigma) + \mO(\Sigma^\dagger W \Sigma)$, we note that
\be \label{eq:kerq21}
\Sigma^\dagger \overline W \Sigma = \mD(\Sigma^\dagger W \Sigma),
\ee
defines the unique projection onto $\mathrm{Ker}(\cQ_\hbar^{(2)})$. Finally, inserting into \eqref{eq:kerq21} the expression
\be
\mD(\Sigma^\dagger W \Sigma) = \frac{1}{2}\tr{W} \un + \frac{1}{2} \tr{\sigma_3 \Sigma^\dagger W \Sigma} \sigma_3,
\ee
and solving for $\overline W$ proves statement \eqref{eq:kernelc2}.

Hence, it was possible to identify the equilibrium spin configuration $\overline W$ for a given two particle interaction potential $\mathbf{V}(r)$, provided that $\cQ_\hbar$ leads to a unique equilibrium distribution in the kernel of $\cQ_\hbar^{(2)}$. This analysis of the quantum collision operator will show to be a crucial ingredient in the following section in order to derive the Bloch as well as the Boltzmann equation. Furthermore, in the above discussion we identified the necessary criteria which allow for the treatment of spin depolarization via a relaxation time operator of the form Eq. \eqref{eq:relaxapprox} together with $\overline W$ given by Eq. \eqref{eq:spindepol}. Such an operator was employed by Possanner {\it et al.} \cite{MeClaudia2011} to derive spin drift-diffusion equations from a semiclassical BE.

However, before we proceed to the semiclassical analysis of the Wigner equation \eqref{eq:wignerhier1} let us briefly discuss the two particular examples already illustrated in Secs. \ref{sec:intkern} and \ref{sec:wigner}.

First, we regard an interaction potential of the form \eqref{eq:vcasea}, i.e. a spin-independent interaction. In this case $\cQ_\hbar^{(2)}$ vanishes for all $W$, i.e. spin polarization is not destroyed. Hence, the quantum collision operator \eqref{eq:dissiwigc1} takes on the form
\be \label{eq:qcasea}
\cQ_\hbar(W) = \int \dint{\eta'} \omega_1(\eta - \eta') |\tilde v(\eta - \eta') |^2 \left [ W(x,\eta',t) - W(x,\eta,t) \right ],
\ee
where we defined
\be
\omega_1(\eta') = 2 \sum_i \rho_i(\eta').
\ee
We note that the collision integral Eq. \eqref{eq:qcasea} independently acts on both spin directions and for each spin species it is of the classical scalar form, however, it has to be kept in mind that $W$ is not a proper distribution function. The weight $\omega_1(\eta')$ is given by the matrix trace of the corrected density-density covariance $\tilde{\mathbf{K}}(\eta')$ (note that the trace is basis independent).

In a similar fashion we obtain for a spin-spin interaction with a potential of the form \eqref{eq:vcaseb} the collision integral
\be
\cQ_\hbar (W) = \int \dint{\eta'} \omega_2(\eta - \eta') [ v(\eta - \eta') ]^2 \left [ \sigma_3 W(x,\eta',t)\sigma_3  - W(x,\eta,t) \right ],
\ee
where
\be
\omega_2(\eta') = 2 (-1)^{\alpha+\alpha'}\tilde \kappa_{\alpha \alpha'\alpha \alpha'}(\eta').
\ee
In particular, according to Eq. \eqref{eq:kernelc2} the projection of $W$ on the kernel of $\cQ_\hbar^{(2)}$ is of the form
\be
\overline W = \frac{1}{2} \tr{W} \un + \frac{1}{2} \tr{\sigma_3 W} \sigma_3,
\ee
i.e. all particles are spin-polarized in $3$-direction in the equilibrium state.

\section{semiclassical limit: spinorial Boltzmann Equation} \label{sec:scalenlimit}

We shall now study the Lindblad Equation \eqref{eq:hier} [or, equivalently, Eq. \eqref{eq:hier1} or the Wigner equation \eqref{eq:wignerhier1}] in a regime in which quantum effects other then spin coherence cease to be observable. The passage from the quantum (or microscopic) world to the classical (or macroscopic) world will be described in terms of a small parameter $\eps$, referred to as the semiclassical parameter or scaled Planck constant. This parameter tends towards zero as quantum effects become more and more negligible.

Before discussing the mathematical subtleties of this transition we shall connect the semiclassical limit to the following physical picture: suppose $\lambda_0$ is a characteristic length on which quantum effects dominate the physics of the system and let $V_0 = \lambda_0^d$ be the volume of a thus defined {\it quantum box} $\cB_0$. Moreover, suppose Eq. \eqref{eq:wignerhier1} provides a proper description of the system in the quantum box $\cB_0$. In the semiclassical picture we are interested in solving Eq. \eqref{eq:wignerhier1} in a domain $\cB_s$ with volume $V_s = V_0 / \eps^d$ for $\eps \ll 1$ with suitable intial and boundary conditions. Hence, $\eps^{-d}$ is a measure of how many quantum boxes with volume $V_0$ fit into the regarded domain $\cB_s$. All wavefunctions are normalized within the domain $\cB_s$. The limit $\eps\to 0$ can be understood as zooming out of the microscopic world $\cB_0$, such that wave-characteristics of particles are on a very small length scale compared to the macroscopic 
domain $\cB_s$, i.e. $V_0 / V_s = \eps^d \to 0 $.

On the other hand, we define a microscopic time scale $\tau_0$ on which quantum effects are dominant and which defines the corresponding {\it quantum time domain} $\cT_0 = [0,\tau_0]$. Then, in the semiclassical limit, we regard a timescale $\tau_s = \tau_0 / \eps$ characterizing the macroscopic time domain $\cT_s$. With the help of $\lambda_0$ and $\tau_0$ we may define the characteristic scales of the {\it microscopic world} $\cB_0 \times \cT_0$ by the relations
\be \label{eq:qscale}
\frac{\epsilon_0 \tau_0}{\hbar} = \frac{\lambda_0 \pi_0}{\hbar} = 1,
\ee
where $\epsilon_0$ and $\pi_0$ stand for the energy scale and the momentum scale associated with the microscopic world $\cB_0 \times \cT_0$, i.e. with Eq. \eqref{eq:wignerhier1}. It is the aim of the following scaling considerations to express Eq. \eqref{eq:wignerhier1} in variables of the {\it macroscopic world} $\cB_s \times \cT_s$.

This is performed in three stages: (i) all functions $f(x,\eta,t)$ which appear in Eq. \eqref{eq:wignerhier1} are rewritten in such a form that we explicitly emphasize their characteristic wavelength and amplitudes. In particular, we rewrite $f(x,\eta,t) = \alpha_c \overline f \left ( \frac{x}{\lambda_c}, \frac{\eta}{\pi_c}, \frac{t}{\tau_c} \right )$, where $\alpha_c$ denotes the characteristic amplitude and $\lambda_c$, $\pi_c$ and $\tau_c$ denote characteristic length-, momentum- and time scales. Thus, the function $\overline f(x, \eta,t)$ is of order one with gradients of order one.

In a second step, we introduce dimensionless variables
\be \label{eq:dimless}
t' = \frac{t}{\tau_s}, \qquad x' = \frac{x}{\lambda_s}, \qquad \eta' = \frac{\eta}{\pi_s} \qquad E' = \frac{E}{\epsilon_s},
\ee
where we introduced the semiclassical scales of the macroscopic world $\cB_s \times \cT_s$. According to the above considerations the semiclassical scales are connected to the quantum scales via
\be
\eps = \frac{\lambda_0}{\lambda_s} = \frac{\tau_0}{\tau_s}.
\ee
It follows from the definition \eqref{eq:qscale} of quantum scales that the above definition implies that
\be \label{eq:epsdef}
\eps = \frac{\hbar}{\epsilon_s \tau_s} = \frac{\hbar}{\lambda_s \pi_s},
\ee
with $\epsilon_s = \epsilon_0$ and $\pi_s = \pi_0$, i.e. the energy and momentum scales remain unaffacted while, by expressing Eq. \eqref{eq:wignerhier1} in the dimensionless variables \eqref{eq:dimless}, we regard a long-time, large-scale limit.

In the third and final step, we define the characteristic lengthscale as well as the characteristic amplitude of all quantities which appear in Eq. \eqref{eq:wignerhier1} by posing suitable scaling assumptions. These assumptions determine the properties of the system under investigation. To be specific we assume in the following semiclassical analysis:
\begin{subequations} \label{eq:scaleassuma}
 \bea
 \mathfrak{h}^{(\eps)}_0(x,\eta) & = & \epsilon_0 \mathfrak{h} \left ( \frac{x}{\lambda_s}, \frac{\eta}{\pi_s} \right ) \un + \eps \epsilon_0 \vec \Omega \left ( \frac{x}{\lambda_s},\frac{\eta}{\pi_s} \right ) \cdot \vec \sigma, \label{eq:assum1}\\
 \mathbf{V}^{(\eps)}(r) & = & a \epsilon_0 \mathbf{V} \left ( \frac{r}{\ell} \right ), \label{eq:assum2} \\
 \mathrm{N}^{(1)}(z) & = & \frac{b}{\lambda_s^d} \mathrm{N}^{(1)}\left ( \frac{z}{\lambda_s} \right ), \label{eq:assum3}
 \eea
\end{subequations}
where we omitted the overlines on the rescaled functions, however, indexed all quantities by $(\eps)$ which implicitely depend on $\eps$. Further, $a,b$ and $\ell$ are scaling parameters which will assume different definite values in the following subsections. Depending on the particular choice of these scaling parameters we shall derive different macroscopic transport models.

Let us briefly interpret the scaling assumptions \eqref{eq:scaleassuma}. The free particle Hamiltonian $\mathfrak{h}_0$ varies on the macroscopic scale $\lambda_s$. We note that the weak scaling of the spin part $\vec \Omega \cdot \vec \sigma$ of the free particle's Hamiltonian $\mathfrak{h}_0$ is necessary in order to preserve spin coherence in the limit $\eps \to 0$, as it was demonstrated by Hajj \cite{raymond}. The particle distribution $\mathrm{N}^{(1)}(z)$ varies on the macroscopic scale $\lambda_s$. The scale on which the interaction potential $\mathbf{V}(r)$  varies is denoted by $\ell$. For instance, for $\ell = \lambda_0$, Eqs. \eqref{eq:scaleassuma} describe short-range interactions between the system's particle and an environment whose density $\mathrm{N}
^{(1)}(z)$ varies on the macroscopic scale $\lambda_s$. On the other hand, for $\ell = \lambda_s$ Eq. \eqref{eq:assum2} accounts for long-range interactions. The parameters $a$ and $b$ are the characteristic amplitudes of $\mathbf{V}(r)$ and $\mathrm{N}^{(1)}(z)$, respectively. For instance, the combination ($a = 1$, $b = \eps$) is referred to as the 'low-density scaling'\cite{spohn77,erdos2005} while the scenario ($a = \eps$, $b = 1$) is denoted the 'weak-coupling scaling'\cite{spohn77,erdos2000}.

In what follows we shall denote the rescaled Wigner distribution matrix $W$ by $W^{(\eps)}$ in order to emphasize its dependence on the semiclassical parameter $\eps$. A short calculation based on the steps illustrated above shows that the rescaled Wigner equation \eqref{eq:wignerhier1} is obtained by replacing $\tau_0 / \hbar^2$ in front of the collision integral as well as $1 / \hbar$ in front of the commutator by $1 / \eps$. This follows from changing to the variables Eq. \eqref{eq:dimless} and employing definition \eqref{eq:epsdef}. In a similar fashion, all $\hbar$-s appearing in Eqs. \eqref{eq:commterm} and \eqref{eq:dissiwig} are replaced by $\eps$. We obtain the rescaled version of Eq. \eqref{eq:wignerhier1} as
\be \label{eq:wignerscale}
\partial_t W^{(\eps)} + \frac{i}{\eps} \cL_\eps \left ( \mathfrak{h}^{mf,(\eps)}_0 \right ) = \frac{1}{\eps} \cQ_\eps \left ( W^{(\eps)} \right ),
\ee
where $\cL_\eps$ and $\cQ_\eps$ denote the rescaled operators \eqref{eq:commterm} and \eqref{eq:dissiwig} and $\mathfrak{h}_0^{mf,(\eps)} = \mathfrak{h}^{(\eps)}_0 + H_{mf}^{(\eps)}$. For $\eps = 1$, the domain on which Eq. \eqref{eq:wignerscale} is defined is equivalent to the microscopic world $\cB_0 \times \cT_0$ while for $\eps \to 0$ we approach the macroscopic world $\cB_s \times \cT_s$.

We assume that a solution $W^{(\eps)}$ of Eq. \eqref{eq:wignerscale}, for suitable intial and boundary conditions, can be written as $W^{(\eps)} = F + \eps W_1 + \ldots$, where $F(x,\eta,t) \in \mH_2(\CC)$ is a positive definite, normalized, slowly varying function of $x$, $\eta$ and $t$, called the spinorial distribution function. This assumption is necessary in order to obtain a well posed semiclassical transport equation because then for $\eps \to 0$, $W^{(\eps)} \to F$ is a proper distribution matrix in the classical sense.

Before we refer to the specific limits, we shall briefly regard the system's particle free Hamiltonian $\mathfrak{h}^{(\eps)}_0$ since this part is independent of the scaling parameters $(a,b,\ell)$. It is demonstrated in App. \ref{app:moyal2} that one obtains
\bea \label{eq:moyal1}
\frac{i}{\eps} \cL_\eps ( \mathfrak{h}^{(\eps)}_0) & = & \frac{i}{\eps} ( \mathfrak{h}^{(\eps)}_0, W^{(\eps)}) \notag \\
&& - \frac{1}{2} \left ( \left \{ \mathfrak{h}^{(\eps)}_0,  W^{(\eps)} \right \}_{x,\eta} - \left \{ W^{(\eps)}, \mathfrak{h}^{(\eps)}_0 \right \}_{x,\eta} \right ) \notag \\
&& + \cO(\eps),
\eea
which is commonly referred to as Moyal's bracket\cite{zachos}. From the scaling assumption \eqref{eq:assum1} we deduce that in zeroth order the spin part $\Omega$ of $\mathfrak{h}^{(\eps)}_0$ appears in the commutator while the scalar part $\mathfrak{h} \un$ enters Poisson's bracket. Hence, provided the semiclassical limit of the mean-field term $\cL_\eps$ and of the quantum collision operator $\cQ_\eps$ exist, we may write Eq. \eqref{eq:wignerscale} in the macroscopic world $\cB_s \times \cT_s$, i.e. for $\eps \to 0$ as
\be \label{eq:befinal}
\partial_t F - \left \{ \mathfrak{h} \un,  F \right \}_{x,\eta} + i [ \vec \Omega \cdot \vec \sigma, F]  + L(F) = Q (F),
\ee
where $L(F)$ denotes the semiclassical limit of the mean-field term and $Q(F)$ the semiclassical limit of the collision integral, see Eq. \eqref{eq:wignerscale}. Please note that the quantity $F$ in Eq. \eqref{eq:befinal} is a proper distribution matrix, i.e. Eq. \eqref{eq:befinal} represents a macroscopic transport equation. It is the aim of the following subsections to specify the particular form of $L(\cdot )$ and $Q(\cdot)$ for different scaling scenarios.

The requirement that $L(F)$ must not diverge, i.e. that Eq. \eqref{eq:befinal} exists, poses constraints on the different combinations of free scaling parameters ($a$, $b$, $\ell$). In particular, inserting Eqs. \eqref{eq:scaleassuma} together with the definition Eq. \eqref{eq:dimless} into the rescaled mean-field interaction \eqref{eq:Hmf3} gives
\bea \label{eq:mean-fieldscale}
H_{mf}^{(\eps)}(x) & = & a b \int \dint{z} n^{(1)}_{\beta \alpha} \left ( z \right )  V_{\alpha \beta} \left [ (x -z )\frac{\lambda_s}{\ell} \right ].
\eea
Let us briefly discuss the two different scenarios for $\ell$ for arbitrary $a,b$: (i) If $\ell= \lambda_s$ the above integral is of order $ab$. (ii) If $\ell = \lambda_0$ the interaction $H_{mf}^{(\eps)}(x)$ is strongly varying in $x$ and $\mathrm{N}^{(1)}(z)$ enters only as constant, i.e.  $\mathrm{N}^{(1)}(0)$. We remark that the contribution of a strongly varying mean-field interaction $H^{(\eps)}_{mf}(x)$ to Eq. \eqref{eq:befinal} vanishes in the semiclassical limit $\eps \to 0$ for most cases. We shall come back to this point in Subsec. \ref{subsec:lowdens1}. Due to the prefactor of $1 / \eps$ in front of $\cL_\eps \left ( H^{(\eps)}_{mf} \right )$ in Eq. \eqref{eq:wignerscale} we require that $ab = \eps$. Particularly interesting are the two specific cases ($a = 1$, $b = \eps$) and ($a = \eps$, $b = 1$), i.e. the low-density and the weak-coupling scaling.

\subsection{Short-range interactions ($\ell = \lambda_0$)} \label{subsec:lowdens1}

Here, we study the semiclassical limit of Eq. \eqref{eq:wignerhier1} under the scaling assumption Eq. \eqref{eq:scaleassuma} for ($a = 1$, $b = \eps$, $\ell = \lambda_0$), i.e. a short-range low density scaling and ($a = \eps$, $b = 1$, $\ell = \lambda_0$), i.e. a short-range weak coupling scaling. It is important to realize that we restrict our discussion to potentials which are integrable in $\RR^d$, i.e. $V_{\alpha \beta}(r) \in L^1(\RR^d)$, such as the  Yukawa potential (screened Coulomb potential). Please note that this excludes the bare Coulomb interaction. However, since in this case $V(r / \eps) = \eps V(r)$, the strongly varying Coulomb interaction is equivalent to the slowly varying Coulomb interaction in the weak coupling scaling and will therefore be discussed in Subsec. \ref{subsec:weakcoupl2}.

First, we note from Eqs. \eqref{eq:mean-fieldscale} that the mean-field contribution vanishes in both scalinges, i.e. $L(F) = 0$, since
\be
H^{(\eps)}_{mf}(x) = \eps \int \dint{z} \int \dint{\xi} n_{\beta \alpha}^{(1)}(\eps z) \tilde{V}_{\alpha \beta}(\xi) \exp ( -i z \cdot \xi )  \exp \left ( \frac{i}{\eps} x\cdot \xi \right ) \to 0,
\ee
due to the Riemann-Lebesgue lemma and $\tilde V_{\alpha \beta}(r) \in L^1(\RR^d)$. Here, the Fourier transform $\tilde{V}_{\alpha \beta}(\xi)$ of $V_{\alpha \beta}(x / \eps)$ is independent of $\eps$, see App. \ref{app:fourier}.

In order to evaluate the semiclassical limit of the collision integral, we make the following hypothesis on the environmental covariance $\mathbf{C}$ defined in Eq. \eqref{eq:defc},
\bea \label{eq:scalec1}
\mathbf{C}^{(\eps)}(z,z') & = & \gamma \mathbf{C}^{(\eps)}(z - z') \notag \\
 & = & \gamma \mathbf{\Gamma}_s \left ( z - z' \right ) + \frac{\gamma}{\eps^{2d}} \mathbf{\Gamma}_0 \left ( \frac{z - z'}{\eps} \right ),
\eea
i.e. we regard a space-homogeneous environment whose covariance has a microscopic and a macroscopic part, $\mathbf{\Gamma}_0$ and $\mathbf{\Gamma}_s$, which are assumed to be behave as $\gamma$ when approaching the macroscopic regime.

In the low density scaling, the resulting collision integral is for $\gamma = \eps$ of the form
\be
Q(F) = Q_0(F),
\ee
where we have (App. \ref{app:dissilimit1})
\bea \label{eq:q01}
Q_0(F) & = &  \int \dint{\eta'} \rho_{i}^{(0)}(\eta - \eta') \left [ S_{i} \left ( \eta -\eta' \right ) F' S_{i} \left (\eta - \eta' \right )  \right. \notag \\
&& - \frac{1}{2} S_{i} \left (\eta - \eta' \right )  S_{i} \left ( \eta -\eta' \right ) F \notag \\
&& \left. -\frac{1}{2} F S_{i} \left ( \eta -\eta' \right ) S_{i} \left (\eta - \eta' \right ) \right ],
\eea
with $F = F(x,\eta,t)$ and $F' = F(x,\eta',t)$. Here, the matrices $S_i(\xi)$ are, again, hermitian linear combinations of the matrices $V_i$ which are obtained by diagonalizing the matrix $\mathbf{\Gamma}_0$, see App. \ref{app:rewrite}. The scalars $\rho_i^{(0)} \in \RR$ are the eigenvalues of $\mathbf{\Gamma}_0$. In the low density scaling, $Q_0$ vanishes for $\gamma = \eps^2$ or weaker while it diverges for $\gamma = 1$ or stronger. The same collision integral \eqref{eq:q01} is obtained in the short-range weak coupling scaling for $\gamma = 1 / \eps$.


We note that the operator \eqref{eq:q01} is a linear collision integral of Boltzmann-form, i.e. in the case of a strongly varying density-density covariance $\mathbf{C}^{(\eps)}$ we obtain the semiclassical Boltzmann equation. The complete discussion of the collision operator can be adapted from Sec. \ref{sec:coll}, however, in the present case it was not necessary to assume that the bath density $\mathrm{N}^{(1)} = \mathrm{const}$. A spinorial Boltzmann equation of such a form has already been postulated by Possanner {\it et al.}\cite{MeClaudia2011}, however, it has not been employed as a basis for the derivation of spin drift-diffusion equations. The collision operator used to derive drift-diffusion equations can be obtained from \eqref{eq:q01} by rewriting the collision operator as sum of an operator accounting for momentum relaxation $Q_1$ and a second operator accounting for spin decoherence $Q_2$, see Sec. \ref{sec:coll}. The latter is then replaced by a relaxation time approximation of the form Eq. \
eqref{eq:relaxapprox} where the projection of $F$ onto the kernel of $Q_2$ has to be assumed to be of the form $\overline F = \frac{1}{2} \tr{F} \un$. Within this work it was possible to identify the necessary criteria allowing for such a treatment, see Sec. \ref{sec:coll}.

\subsection{Long-range interactions ($\ell = \lambda_s$)} \label{subsec:weakcoupl2}

Let us regard an interaction potential varying on the macroscopic scale, i.e. $\ell = \lambda_s$ in Eq. \eqref{eq:scaleassuma}. Again, we shall regard the low density ($a = 1$, $b = \eps$, $\ell = \lambda_s$) as well as the weak coupling ($a = \eps$, $b =1$, $\ell =\lambda_s$) case. Please note that the strongly varying Coulomb potential is equivalent to the slowly varying Coulomb potential in the weak coupling scaling. From Eq. \eqref{eq:mean-fieldscale} we deduce for both scalings
\be
L(F) = i [ H_{mf} , F],
\ee
where $H_{mf}$ is the macroscopic mean-field interaction of the form \eqref{eq:Hmf3}, i.e. independent of $\eps$.

In the case of long range interactions we employ the following ansatz for the environmental covariance $\mathbf{C}$:
\bea \label{eq:scalec2}
\mathbf{C}^{(\eps)}(z,z') & = & \gamma \mathbf{C}^{(\eps)}(z,z') \notag \\
 & = & \gamma \mathbf{\Gamma}_s (z,z') + \frac{\gamma}{\eps^{2d}} \mathbf{\Gamma}_0 \left ( \frac{z}{\eps} , \frac{z'}{\eps} \right ),
\eea
i.e. in contrast to Eq. \eqref{eq:scalec1} we do not restrict to the space homogeneous case.

In the low density scaling, the collision operator $Q(F)$ takes on the form (App. \ref{app:dissilimit2})
\be 
Q(F) = Q_{LD}(F) = Q_s(F) + Q_n(F),
\ee
where for $\gamma = \eps$
\begin{subequations}\label{eq:qs}
\bea \label{eq:qs1}
Q_s(F) & = & \frac{2}{(2 \pi)^d}\int \dint{z} \dint{z'} \gamma^{(s)}_{\beta \beta' \alpha \alpha'}(z, z') \notag \\
&& \times \Bigl [ V_{\alpha \beta} \left ( x  - z \right ) F V_{\alpha'\beta' } \left ( x  - z' \right )  \notag \\
&& - \frac{1}{2} V_{\alpha\beta } \left ( x  - z \right ) V_{\alpha'\beta' } \left ( x - z' \right ) F \notag \\
&& \left. -\frac{1}{2} F V_{\alpha\beta } \left ( x  - z \right )V_{\alpha'\beta' } \left ( x -  z' \right ) \right ],
\eea
stems from the slowly varying part of the environmental covariance and
\bea \label{eq:qn}
Q_n(F) & = & \frac{2}{(2 \pi)^d}\int \dint{z} n^{(1)}_{\beta \alpha}(z) \notag \\
&& \times \Bigl [ V_{\alpha \beta} \left ( x  - z \right ) F V_{\alpha\beta } \left ( x -z \right )  \notag \\
&& - \frac{1}{2} V_{\alpha\beta } \left ( x -z  \right ) V_{\alpha'\beta' } \left ( x -z \right ) F \notag \\
&& \left. -\frac{1}{2} F V_{\alpha\beta } \left ( x -z \right )V_{\alpha'\beta' } \left ( x -z \right ) \right ],
\eea
\end{subequations}
stems from the matrix $\mathbf{D}^{(\eps)}$. In the long-range weak coupling limit $Q(F)$ reads for $\gamma = 1 / \eps$ (App. \ref{app:dissilimit2})
\be
Q(F) = Q_{WC}(F) =  Q_s(F).
\ee
Please note that we refrained from writing Eqs. \eqref{eq:qs1} and \eqref{eq:qn} as momentum space integrals and in the basis described in App. \ref{app:rewrite} for the sake of a more transparent notation. Of course, we may obtain such a representation by replacing the interaction potential by its (macroscopic) Fourier representation and carrying out the steps in App. \ref{app:rewrite}. (In particular, for Eq. \eqref{eq:qs1} we note that in the case that the environmental covariance obeys $\mathbf{\Gamma_s}(z,z') = \mathbf{\Gamma}_s(z - z')$ the representation as a momentum space integral will turn out to be more convenient. Moreover, the form of $Q_s(F)$ will in this case be very similar to Eq. \eqref{eq:q01}, i.e. one obtains a Boltzmann collision integral.)

Let us briefly discuss the ensuing transport equation in the long-range weak coupling limit for $\gamma = 1 / \eps$. According to the above discussion we have
\be \label{eq:trans}
\partial_t F - \left \{ \mathfrak{h} \un,  F \right \}_{x,\eta} + i [ \vec \Omega \cdot \vec \sigma + H_{mf}, F]  =  Q_s (F),
\ee
The transport equation of the spin part $\vec f$ of $F$ is of the form
\bea \label{eq:bloch}
\partial_t \vec f & + \left ( \nabla_x \mathfrak{h} \cdot \nabla_\eta \vec f - \nabla_\eta \mathfrak{h} \cdot \nabla_x \vec f \right )  + 2 \vec f \times \left ( \vec \Omega + \vec h_{mf} \right ) \notag \\
 & = - \int \dint{z} \dint{z'} \gamma_i^{(s)}(z,z')  \vec s_i(x,z,z') \times \vec s_i(x,z,z') \times \vec f ,
\eea
where we employed Eqs. \eqref{eq:paulibase} and \eqref{eq:commutpauli} and rewrote $Q_s$ with the help of the steps outlined in App. \ref{app:rewrite}. In particular, the vectors $\vec s_i(x,z,z')$ are linear combinations of the spin parts $\vec v_i(x - z)$ of $V_i(x-z)$, see Eq. \eqref{eq:vpauli}, however, the weights may be functions of $z$ and $z'$. The quantities $\gamma_i^{(s)}(z,z')$ are the eigenvalues of $\mathbf{\Gamma}_s(z,z')$. Equation \eqref{eq:bloch} describes the precession of the directional spin polarization $\vec f$ under the influence of an external field $\vec \Omega$ and an interaction with the environment. This interaction induces an additional field $\vec h_{mf}$ and gives rise to the dissipator $Q_s$, which relaxes the vector $\vec f$ into a predefined direction, see Sec. \ref{sec:coll}. 

Let us consider a spin located at a certain lattice point ({\em 'system'}) which interacts with other spins located at different lattices sites ({\em 'environment'}). Since the system's spin cannot move on the lattice the scalar part of the free particle Hamiltonian vanishes, $\mathfrak{h} = 0$, i.e. it has no kinetic part and no scalar field is externally applied. We recognize that in this particular case Eq. \eqref{eq:bloch} gives rise to the macroscopic Bloch equations of magnetism. To be more specific, we replace in Eq. \eqref{eq:bloch} the dissipator $Q_s$ by a relaxation time ansatz as discussed in Sec. \ref{sec:coll} and interpret the spin part $\vec f = 1/ 2 \tr{F \vec \sigma}$ as the magnetization of the system.

We emphasize that it was, therefore, possible to derive the Boltzmann as well as the Bloch equations from the spinorial Wigner equation derived within this work. In a similar fashion one may impose further scaling assumptions and study the resulting classical transport models as well as their quantum corrections.

\section{Summary} \label{sec:summary}

We derived a linear quantum collision operator for the spinorial Wigner equation. Furthermore, it was demonstrated that the Wigner equation gives rise to several linear semiclassical spin-transport models. We derived the Bloch equations as well as the linear Boltzmann equation as an example. Let us briefly summarize the main aspects of the derivation.

We investigated the dissipative dynamics of a spin-$1 / 2$ quantum particle, referred to as the system, in contact with its environment, which is ni thermal equilibrium. It has been shown by Possanner and Stickler \cite{possanner12} that in the limit of vanishing system-environment correlations these dynamics are properly described by the Lindblad equation \eqref{eq:hier}. The latter served as a basis of the current study. The Wigner representation of the Lindblad equation \eqref{eq:hier} is a spinorial Wigner equation \eqref{eq:wignerhier1} equipped with a quantum collision operator \eqref{eq:quantcoll}. It is then demonstrated that the latter can be cast into the form of a Boltzmann collision integral, Eq. \eqref{eq:dissiwig2}, provided that the spinorial density of the environment is constant and that the spinorial density-density covariance of the environment is space homogeneous. The $2 \times 2$ hermitian scattering matrices $S_i$ are uniquely determined by the spin-dependent two-particle interaction 
potential $\mathbf{V}(r)$ and by the modified spinorial density-density covariance $\mathbf{K}(z,z')$ of the environment, defined in Eq. \eqref{eq:kappadef}. The eigenvalues of the scattering matrices $S_i$ are the scattering rates for the two spin species.

Moreover, the quantum collision operator \eqref{eq:dissiwig2} is composed of two qualitatively rather different parts. The first part changes solely the spinorial momentum density of the system while the second part accounts for local spin flip processes. Hence, the second part modifies the local spin polarization of the system in a fashion uniquely determined by the scattering matrices $S_i$ and the eigenvalues of the modified spinorial density-density covariance $\mathbf{K}(z,z')$. Furthermore, it is possible to identify clear criteria under which the interaction between the system's particle and the environment leads to spin decoherence or even spin depolarization in the long time limit.

Finally, we performed a semiclassical analysis of the spinorial Wigner equation \eqref{eq:wignerhier1}, i.e. we regarded the dynamics of the system's particle in a regime in which quantum effects other than spin-coherence cease to be observable. We restricted our discussion to the well established low-density and weak-coupling limits. In principle, several semiclassical evolution equations for a positive definite, hermitian distribution matrix $F$ can be obtained. As two particularly interesting examples we note the derivation of the Bloch equations for long range interactions and the spinorial Boltzmann equation for short range interactions and a spatially homogeneous environmental density-density covariance.\footnote{It is interesting to note that the covariance in the collision integral is a direct consequence of the requirement of well-posedness of the hierarchy of master equations on the quantum scale\cite{possanner12}.} This form of the Boltzmann equation has already been used for deriving spin-
coherent drift-diffusion equations in magnetic multilayers\cite{MeClaudia2011}.

In summary, we remark that within this work it was possible to systematically establish the link between a full quantum-mechanical treatment of a composite quantum system by means of the von-Neumann equation and macroscopic linear spin-transport models featuring dissipation such as the spin drift-diffusion models. This makes the derived equations particularly interesting for applications involving graphene\cite{morandi} (pseudo-spin formalism) or magnetically doped semiconductors\cite{fabian1,fabian2}.

possible to augment the resulting transport models with corrections which result from the Born-Markov limit as well as from the semiclassical limit if deemed necessary. Within this work we restricted to the zeroth order equations in both scalings but the evaluation of higher order corrections is straight-forward.

\begin{acknowledgments}
The authors are very grateful to E. Schachinger for carefully reading the manuscript. B.A.S. was supported by the Austrian Science Fund (FWF): P221290-N16.
\end{acknowledgments}

\appendix

\section{Derivation of Eq. \eqref{eq:dissi2}} \label{app:dissi1}

We evaluate the matrix elements of the dissipator \eqref{eq:dissimed} by investigating the terms containing $\trB{\hat H_I \hat H_I \hat \un \otimes \hat \chi_B}$. In what follows we employ Einstein's summation convention in order ot simplify the notation. We obtain
\bea \label{lololo}
 \trB{\hat H_I \hat H_I \hat \un \otimes \hat \chi_B}(x,y) & = & \dinto{z}{z'} \left [ N (N-1) \chi^{(2)}_{ \beta\beta'\alpha\alpha' }(z,z') + N \chi^{(1)}_{ \beta\alpha }(z) \delta_{\alpha \alpha'} \delta_{\beta \beta'} \delta(z - z' ) \right ] \notag \\
 &&\times V_{\alpha \beta}(x-z)V_{\alpha' \beta'}(y-z')  \delta(x - y)
\eea
where $\chi^{(2)}_{ \beta \beta'\alpha\alpha' }(z,z')\in\CC$ stems from Eq. \eqref{eq:chi2} for identical particles (indices $n,m$ omitted). In a similar fashion one obtains the following relations
\begin{subequations} \label{lalala}
\bea
 \trB{\hat H_I \hat \rho\otimes \hat \chi_B \hat H_I}(x,y) & = & N ( N -1) \dinto{z}{z'}  \chi^{(2)}_{ \beta\beta'\alpha\alpha' }(z,z') V_{\alpha \beta}(x-z)\rho(x,y)V_{\alpha'\beta'}(y-z') \notag \\
  && + N \int \dint{z}  \chi^{(1)}_{ \beta\alpha }(z) V_{\alpha \beta}(x-z)\rho(x,y)V_{\alpha\beta}(y-z),
\eea
\bea
\trB{\hat H_I^2 \hat \rho \otimes \hat \chi_B}(x,y) & = & N ( N-1) \dinto{z}{z'} \chi^{(2)}_{ \beta\beta'\alpha\alpha' }(z,z') V_{\alpha \beta}(x-z)V_{\alpha' \beta'}(x-z') \rho(x,y) \notag \\
 && + N \int \dint{z} \chi^{(1)}_{ \beta\alpha }(z) V_{\alpha \beta}(x-z)V_{\alpha \beta}(x-z) \rho(x,y),
\eea
\bea
\trB{\hat \rho \otimes \hat \chi_B \hat H_I^2}(x,y) & = & N ( N-1) \dinto{z}{z'} \chi^{(2)}_{ \beta\beta'\alpha\alpha' }(z,z') \rho(x,y) V_{\alpha \beta}(y-z)V_{\alpha' \beta'}(y-z') \notag \\
 && + N \int \dint{z} \chi^{(1)}_{ \beta\alpha }(z) \rho(x,y) V_{\alpha \beta}(y-z)V_{\alpha \beta}(y-z) ,
\eea
\be
\left ( \hat H_{mf} \hat \rho \hat H_{mf} \right )(x,y) = N^2 \sinto{z} \dint{z'}  \chi^{(1)}_{ \beta\alpha }(z) \chi^{(1)}_{ \beta' \alpha' }(z') V_{\alpha \beta}(x-z) \rho(x,y) V_{\alpha' \beta'}( y - z'  ),
\ee
\be
\left ( \hat H_{mf}^2 \hat \rho \right )(x,y) = N^2 \sinto{z} \dint{z'} \chi^{(1)}_{ \beta\alpha }(z) \chi^{(1)}_{ \beta' \alpha' }(z') V_{\alpha \beta}(x-z) V_{\alpha' \beta'}( x - z'  ) \rho(x,y),
\ee
and
\be
\left (\hat \rho \hat H_{mf}^2  \right )(x,y) = N^2 \sinto{z} \dint{z'} \chi^{(1)}_{ \beta\alpha }(z) \chi^{(1)}_{ \beta' \alpha' }(z') \rho(x,y) V_{\alpha \beta}(y-z) V_{\alpha' \beta'}( y - z'  ).
\ee
\end{subequations}
Using that $N(N-1) \approx N^2$, applying definitions \eqref{replace1} together with \eqref{eq:kappadef} and \eqref{eq:defc} and inserting the relations \eqref{lalala} into Eq. \eqref{eq:dissimed} yields the final result Eq. \eqref{eq:dissi2}.

\section{Derivation of the Moyal bracket} \label{app:moyal}

Within this appendix we derive the phase space symbol of the commutator Eq. \eqref{eq:commterm}, i.e. the Moyal bracket. We calculate the Wigner transform of the element $\matel{x}{\hat A \hat \rho}{y} = \left (\hat A \hat \rho \right )(x,y)$ for some operator $\hat A$ as
\bea \label{eq:moyal2}
\cW \left [ \left ( \hat A \hat \rho \right )(x,y)\right ] & = & \frac{1}{(2 \pi \hbar)^d}\int \dint{x'} \dint{z} A \left ( x+\frac{1}{2} x', z \right ) \rho \left (z,x - \frac{1}{2}x' \right ) \exp ( - i x' \cdot \eta ) \notag \\
& = & \frac{1}{(2 \pi \hbar)^{2d}} \int \dint{x'} \dint{z} \int \dint{\xi} \dint{\xi'} \mathfrak{a} \left ( \frac{x+z}{2} + \frac{1}{4} x', \xi \right ) \notag \\
&& \times W \left ( \frac{z + x}{2} - \frac{1}{4}x',  \xi' \right ) \exp \left [ \frac{i}{\hbar} \xi \cdot \left ( x + \frac{1}{2} x' - z \right ) \right ] \notag \\
&& \times \exp \left [ \frac{i}{\hbar} \xi' \cdot \left ( z - x + \frac{1}{2} x' \right ) \right ]  \exp \left ( - \frac{i}{\hbar} x' \cdot \eta \right ),
\eea
where we used Eq. \eqref{invwig} and $\matel{x}{\hat A}{y} = A(x,y) = ( 2 \pi \hbar)^{-d} \cW^{-1}[\mathfrak{a}(x,\eta)]$. We rearrange the exponential terms as
\be
  \frac{i}{\hbar} x \cdot (\xi - \xi') + \frac{i}{\hbar} z \cdot ( \xi' -\xi ) - \frac{i}{\hbar} x' \cdot \left [ \eta - \frac{1}{2} ( \xi + \xi' ) \right ] \notag
\ee
which suggests the substitution $\tilde \xi = \xi - \xi'$ and $\overline \xi = \frac{1}{2} ( \xi + \xi' )$. Hence, we have
\bea
\cW \left [ \left ( \hat A \hat \rho \right )(x,y)\right ] & = & \frac{1}{(2 \pi \hbar)^{2d}} \int \dint{x'} \dint{z} \int \dint{\overline \xi} \dint{\tilde \xi} \mathfrak{a} \left ( \frac{x+z}{2} + \frac{1}{4} x', \overline \xi + \frac{1}{2} \tilde \xi \right ) \notag \\
&& \times W \left ( \frac{x + z}{2} - \frac{1}{4}x',  \overline \xi - \frac{1}{2} \tilde \xi \right ) \exp \left [ \frac{i}{\hbar} (x - z) \cdot \tilde \xi +\frac{i}{\hbar}x' \cdot (\overline \xi - \eta) \right ].
\eea
This expression may be rewritten in a more convenient form by replacing $z' = \frac{x + z}{2}$ and also $\hat \xi = 2 \tilde \xi$. Finally, we obtain the first term of Eq. \eqref{eq:commterm} as
\bea \label{eq:moyalmed}
 \cW \left [ \left ( \hat A \hat \rho \right )(x,y)\right ] & = & \frac{1}{(2 \pi \hbar)^{2d}} \int \dint{x'} \dint{z'} \int \dint{\overline \xi} \dint{\hat \xi} \mathfrak{a} \left ( z' + \frac{1}{4} x', \overline \xi + \frac{1}{4} \hat \xi \right ) \notag \\
&& \times W \left ( z' - \frac{1}{4}x',  \overline \xi - \frac{1}{4} \hat \xi \right ) \exp \left [ \frac{i}{\hbar} (x - z') \cdot \hat \xi +\frac{i}{\hbar}x' \cdot (\overline \xi - \eta) \right ].
\eea
A similar calculation for the term $\matel{x}{\hat \rho \hat A}{y} = \left ( \hat \rho \hat A \right )(x,y)$ gives
\bea
 \cW \left [ \left ( \hat \rho \hat A \right )(x,y)\right ] & = & \frac{1}{(2 \pi \hbar)^{2d}} \int \dint{x'} \dint{z'} \int \dint{\overline \xi} \dint{\hat \xi} W \left ( z' + \frac{1}{4} x', \overline \xi + \frac{1}{4} \hat \xi \right ) \notag \\
&& \times \mathfrak{a} \left ( z' - \frac{1}{4}x',  \overline \xi - \frac{1}{4} \hat \xi \right ) \exp \left [ \frac{i}{\hbar} (x - z') \cdot \hat \xi +\frac{i}{\hbar}x' \cdot (\overline \xi - \eta) \right ].
\eea
Combining this relation with Eq. \eqref{eq:moyalmed} gives Eq. \eqref{eq:commterm} as
\bea
\cW \left [ \cL(A)(x,y) \right ] (x, \eta) & = & \frac{1}{(2 \pi \hbar)^{2d}} \int \dint{z} \dint{z'} \int \dint{\xi} \dint{\xi'} \left [ \mathfrak{a} \left ( z' + \frac{1}{4} z, \xi + \frac{1}{4} \xi' \right ) \right. \notag \\
&& \left.  W \left ( z' - \frac{1}{4}z, \xi - \frac{1}{4} \xi',t \right ) - W \left ( z' + \frac{1}{4} z, \xi + \frac{1}{4} \xi',t \right ) \right. \notag \\
&& \left.  \mathfrak{a} \left ( z' - \frac{1}{4}z, \xi - \frac{1}{4} \xi' \right ) \right ] \exp \left [  \frac{i}{\hbar} \xi' \cdot (x - z') \right ] \notag \\
&& \times \exp \left [ - \frac{i}{\hbar} z \cdot \left ( \eta - \xi \right ) \right ].
\eea

We shall now investigate the particular case $\matel{x}{\hat B}{y} = B(x,y) = B(x) \delta(x - y)$ where it follows from Eq. \eqref{def:wignerphys} that $\mathfrak{b}(x,\eta) \equiv B(x)$. We retur to Eq. \eqref{eq:moyal2} in order to obtain
\bea
\cW \left [ \left ( \hat B \hat \rho \right )(x,y)\right ] & = & \frac{1}{(2 \pi \hbar)^d}\int \dint{x'} B \left ( x+\frac{1}{2} x' \right ) \rho \left (x + \frac{1}{2} x',x - \frac{1}{2}x' \right ) \exp ( - i x' \cdot \eta ) \notag \\
& = & \frac{1}{(2 \pi \hbar)^{d}} \int \dint{x'} \int \dint{\xi} B \left ( x + \frac{1}{2} x' \right ) W \left ( x,  \xi \right ) \exp \left [ \frac{i}{\hbar} (\xi - \eta) \cdot x' \right ] ,
\eea
and, therefore, for $B(x) = b(x) \un$ the well known result Eq. \eqref{eq:vlasov2}
\bea
\cW \left [ \cL \left ( b \un \right ) (x,y)\right ] & = & \frac{1}{(2 \pi \hbar)^d}\int \dint{x'} \int \dint{\xi'} \left [ b \left ( x+\frac{1}{2} x' \right ) -  b \left ( x-\frac{1}{2} x' \right ) \right ] W' \notag \\ 
 && \times \exp \left [ -\frac{i}{\hbar} x' \cdot (\eta - \xi') \right ] ,
\eea
follows, where we defined $W' = W(x,\xi',t)$.

\section{The Fourier transform} \label{app:fourier}

We define the Fourier transform of an operator, which is diagonal in position space, i.e. $V(x,y) = V(x) \delta(x - y)$ as
\be \label{eq:fourierV2}
\tilde V(\eta) = \frac{1}{(2 \pi \hbar)^d} \int \dint x V(x) \exp \left ( - \frac{i}{\hbar} x \cdot \eta \right ),
\ee
and, therefore, its inverse as
\be
V(x) = \int \dint \eta \tilde V(\eta) \exp \left (  \frac{i}{\hbar} x \cdot \eta \right ).
\ee

In a similar fashion we define the Fourier transform of a function $K(x,y)$ which stems from a two particle operator i.e. $K(x,y,x',y') = K(x,y) \delta(x-x') \delta(y - y')$ as
\be \label{eq:fourierkappa}
\tilde K(\xi,\xi') = \frac{1}{(2 \pi \hbar)^{2d}} \int \dint x \dint y K(x,y) \exp \left [ - \frac{i}{\hbar} \left ( x \cdot \xi + y \cdot \xi' \right ) \right ].
\ee
In particular, if $K(x,y) = K(x - y)$ we obtain
\bea
\tilde K(\xi, \xi') & = & \frac{1}{(2 \pi \hbar)^{d}}\int \dint x K(x) \exp \left ( -\frac{i}{\hbar} x \cdot \xi\right )  \delta(\xi + \xi') \notag \\
 & = & \tilde K \left ( \frac{\xi - \xi'}{2} \right ) \delta(\xi + \xi').
\eea

Moreover, we will frequently employ the identities
\be \label{eq:dirac1}
\delta (x) = \frac{1}{(2 \pi)^d} \int \dint \eta \exp ( i \eta \cdot x),
\ee
and
\be \label{eq:dirac2}
\delta (\eta) = \frac{1}{(2 \pi)^d}\int \dint x \exp ( - i \eta \cdot x),
\ee
where $\delta ( \cdot )$ denotes Dirac's delta distribution.

The semiclassical Fourier transform is obtained by replacing all appearing $\hbar$-s in Eq. \eqref{eq:fourierV2} by $\eps$. Hence, we have
\be
\tilde V^{(\eps)}(\eta) = \frac{1}{(2 \pi \eps)^d} \int \dint x V^{(\eps)}(x) \exp \left ( - \frac{i}{\eps} x \cdot \eta \right ),
\ee
and, therefore, also
\be
V^{(\eps)}(x) = \int \dint \eta \tilde V^{(\eps)}(\eta) \exp \left (  \frac{i}{\eps} x \cdot \eta \right ),
\ee
for a single particle operator which is diagonal in position space. Here, the index $(\eps)$ signifies that $V$ may still be a function of $\eps$. In the particular case that $V^{(\eps)}(x) = V( x / \eps )$ we obtain the important result that
\be\label{eq:vinv}
\tilde V^{(\eps)}(\eta) =  \frac{1}{(2 \pi \eps)^d} \int \dint x V \left ( \frac{x}{\eps} \right ) \exp \left ( - \frac{i}{\eps} x \cdot \eta \right ) = \tilde V(\eta),
\ee
is independent of $\eps$. In a similar fashion we obtain that
\be
\tilde K^{(\eps)}(\xi,\xi') = \frac{1}{(2 \pi \eps)^{2d}}\int \dint x \dint y K^{(\eps)} \left (x,y \right ) \exp \left [ -\frac{i}{\eps} \left ( x \cdot \xi + y \cdot \xi' \right ) \right ],
\ee
is independent of $\eps$ for $K^{(\eps)}(x,y)$ strongly varying. In particular,
\be \label{eq:kinv}
\tilde K^{(\eps)}(\xi,\xi') = \frac{1}{(2 \pi \eps)^{2d}}\int \dint x \dint y K \left ( \frac{x}{\eps} ,\frac{y}{\eps} \right ) \exp \left [ -\frac{i}{\eps} \left ( x \cdot \xi + y \cdot \xi' \right ) \right ] = \tilde K(\xi,\xi').
\ee

\section{Rewriting the dissipator}\label{app:rewrite}

We shall derive the representation \eqref{eq:dissiwig3} of the dissipator. In what follows we shall drop the explicit notation of the momentum argument $\xi$ as well as the tildes in order to simplify the expressions.
We note that we can express the matrix $\mathbf{K}$ with the help of the Pauli base \eqref{eq:paulibase} as
\be
\mathbf{K} = \mathrm{K}_0 \otimes \un + \vec{ \mathrm{K}} \odot \vec \sigma,
\ee
where $\vec{\mathrm{K}} \odot \vec \sigma = \mathrm{K}_i \otimes \sigma_i$ where $i$ runs from $1$ to $3$ and we defined the components $\mathrm{K}_i \in \mH_2(\CC)$ for $i = 0,\ldots,3$ according to Eq. \eqref{eq:paulibase}. We express the remaining hermitian matrices $\mathrm{K}_i$ in a similar fashion in order to obtain
\be
\mathrm{K}_i = k_i \un + \vec k_i \cdot \vec \sigma = k_{i0} \un + k_{ij} \sigma_j,
\ee
and, hence,
\be
\mathbf{K} = k_{00} \un \otimes \un + k_{0j} \sigma_j \otimes \un + k_{i0} \un \otimes \sigma_i + k_{ij} \sigma_j \otimes \sigma_i.
\ee
It follows from the definition \eqref{eq:paulibase} that $k_{ij} \in \RR$. Furthermore, from the indistinguishability of bath particles we note that $\kappa_{\alpha \alpha' \beta \beta'} = \kappa_{\alpha' \alpha \beta' \beta}$ and, therefore,
\be \label{eq:sym1}
k_{ij} = k_{ji} \qquad i,j = 0,\ldots,3.
\ee
In a similar fashion we decompose the interaction potential $\mathbf{V} \in \mH_2(\CC) \otimes \mH_2(\CC)$ with respect to the second particle with the help of Eq. \eqref{eq:paulibase} as Eq. \eqref{eq:vpauli} where $V_i \in \mH_2(\CC)$ are given by Eqs. \eqref{eq:vpauli}. With the help of these definitions we rewrite a typical sum which appears in Eq. \eqref{eq:dissiwig2} as
\be\label{eq:sum1}
\kappa_{\beta \beta' \alpha \alpha'} V_{\alpha \beta} V_{\alpha' \beta'} = k_{ij} V_i V_j,
\ee
where the sum goes over all $i,j = 0,\ldots,3$. We note that we can understand the above sum Eq. \eqref{eq:sum1} as the scalar product between a vector $\vec \cV = (V_0,V_1,V_2,V_3)^T \in \RR^4 \otimes \mH_2(\CC)$ and the rotated vector $\mathbf{\cK} \cV$, where the matrix $\mathbf{\cK} = \{ k_{ij} \} \in \RR^{4 \times 4}$ is symmetric due to Eq. \eqref{eq:sym1}, i.e. $\mathbf{\cK}^T = \mathbf{\cK}$. Since all elements of $\mathbf{\cK}$ are real and since $\mathbf{\cK}$ is symmetric, it follows that the $\mathbf{\cK}$ may be diagonalized by an orthogonal matrix $\mathbf{\cU} \in \RR^{4 \times 4}$, where $\mathbf{\cU}^T = \mathbf{\cU}^{-1}$. Hence, denoting by $(\cdot, \cdot )$ the scalar product in $\RR^4$, we have
\be
k_{ij} V_i V_j = \left ( \cV, \mathbf{\cK} \cV \right ) = (\cV, \mathbf{\cU}^T \cR \mathbf{\cU} \cV ) = (\mathbf{\cU} \cV, \cR \mathbf{\cU} \cV) = (\cS, \cR \cS),
\ee
where we defined $\cS = \mathbf{\cU} \cV \in \RR^4 \otimes \mH_2(\CC)$. Denoting by $S_i$ the components of the vector $\cS$, and by $\rho_i$ the diagonal elements of $\cR$ where $S_i \in \mH_2(\CC)$ and $\rho_i \in \RR$, we obtain
\be
\kappa_{\beta \beta' \alpha \alpha'} V_{\alpha \beta} V_{\alpha' \beta'} = \rho_i S_i S_i.
\ee
We remark that this convenient form of the sum Eq. \eqref{eq:sum1} is a result of the indistinguishability of bath particles, which assures that $\mathbf{K}$ is a real, symmetric matrix.

\section{The kernel of $\cQ_\hbar^{(2)}$} \label{app:dissicomm}

We shall briefly demonstrate that $\cQ_\hbar^{(2)}(A) = 0$ is equivalent to $[S_i(\xi) , A] = 0$ for all $\xi$ and all $i$ if detailed balance is required for $A$, where $A, S_i(\xi) \in \mH_2(\CC)$. Detailed balance means that each term contributing to $\cQ_\hbar^{(2)}$, Eq. \eqref{eq:dissiwigc2}, vanishes individually. Hence, it suffices to investigate the operator
\be
\tilde{\cQ}_\hbar^{(2)}(A) = \frac{1}{2} \int \dint{\xi} [[ \Lambda(\xi), A], \Lambda(\xi)],
\ee
where $\Lambda(\xi) \in \mH_2(\CC)$. We assume that $A \neq \un$ however $\tilde{\cQ}_\hbar^{(2)}(A) = 0$. Then, clearly
\be
\mathrm{tr} \left [\tilde{\cQ}_\hbar^{(2)}(A) A \right ] = \int \dint{\xi} \mathrm{tr} \left [ \Lambda(\xi) A \Lambda(\xi) A - \Lambda(\xi)\Lambda(\xi) A^2 \right ] = 0.
\ee
We rewrite the first term of this equation for all $\xi$ as
\be
\mathrm{tr} \left [\Lambda(\xi) A \Lambda(\xi) A \right ] = \lambda_{ii}(\xi) \lambda_{kk}(\xi) \vert a_{ik}(\xi) \vert^2,
\ee
where the $\lambda_{ii}$ are the eigenvalues of $\Lambda(\xi)$ and the $a_{ik}(\xi)$ are matrix elements of $A$ represented in the eigenbasis of $\Lambda(\xi)$. In a similar fashion, we obtain for all $\xi$
\be
\mathrm{tr} \left [\Lambda(\xi) \Lambda(\xi)  A^2 \right ] = \lambda_{ii}^2(\xi) \vert  a_{ik}(\xi) \vert^2.
\ee
Since $\Lambda(\xi) \in \mH_2(\CC)$,
\be
\mathrm{tr} \left [ \Lambda(\xi) A \Lambda(\xi) A - \Lambda(\xi)\Lambda(\xi) A^2 \right ] = - \left [ \lambda_{11} ( \xi) - \lambda_{22}(\xi) \right ]^2 \vert a_{12}(\xi) \vert^2 \leq 0,
\ee
and, therefore,
\be
\tilde{\cQ}_\hbar^{(2)}(A) \leq 0, \qquad \tilde{\cQ}_\hbar^{(2)}(A) = 0 \Rightarrow [\Lambda(\xi), A] = 0 \quad \forall \xi.
\ee
The statement $[\Lambda(\xi),A] = 0 \Rightarrow \tilde{\cQ}_\hbar^{(2)}(A) = 0$ is trivial and, therefore,
\be
\cQ_\hbar^{(2)}(A) = 0 \Leftrightarrow [S_i(\xi),A ] = 0 \quad \forall \xi \text{ and } \forall i,
\ee
if detailed balance is required for $A$.

\section{Derivation of the Moyal product} \label{app:moyal2}

In this Appendix we briefly present the derivation of Eq. \eqref{eq:moyal1}, i.e. the Moyal product, where the Moyal bracket has been specified in App. \ref{app:moyal}. We start with Eq. \eqref{eq:moyalmed}. We rewrite this equation in dimensionless variables, see Eq. \eqref{eq:dimless}, in order to obtain
\bea \label{eq:moyal3}
  \cW_\eps \left [ \left ( \hat A \hat \rho \right )(x,y)\right ] & = & \frac{1}{(2 \pi \eps)^{2d}} \int \dint{x'} \dint{z'} \int \dint{\overline \xi} \dint{\hat \xi} \mathfrak{a}^{(\eps)} \left ( z' + \frac{1}{4} x', \overline \xi + \frac{1}{4} \hat \xi \right ) \notag \\
&& \times W^{(\eps)} \left ( z' - \frac{1}{4}x',  \overline \xi - \frac{1}{4} \hat \xi \right ) \exp \left [ \frac{i}{\eps} (x - z') \cdot \hat \xi +\frac{i}{\eps}x' \cdot (\overline \xi - \eta) \right ] \notag \\
 & = & \frac{1}{(2 \pi)^{2d}}\int \dint{x'} \dint{z'} \int \dint{\overline \xi} \dint{\hat \xi} \mathfrak{a}^{(\eps)} \left ( z' + \frac{\eps}{4} x', \overline \xi + \frac{\eps}{4} \hat \xi \right ) \notag \\
&& \times W^{(\eps)} \left ( z' - \frac{\eps}{4}x',  \overline \xi - \frac{\eps}{4} \hat \xi \right ) \exp \left [ i (x - z') \cdot \hat \xi +i x' \cdot (\overline \xi - \eta) \right ]
\eea
The expression \eqref{eq:moyal3} can now be expanded in a Taylor series in terms of $\eps$ around $\eps = 0$ under the assumption that $\mathfrak{a}^{(\eps)}$ and $W^{(\eps)}$ are both slowly varying functions with amplitudes of order one. Then, we obtain in zeroth order
\bea \label{eq:moyal4}
\cO_0 & = & \mathfrak{a}^{(0)}  W^{(0)} ,
\eea
where Eqs. \eqref{eq:dirac1} and \eqref{eq:dirac2} have been used in order to eliminate of the integrals in Eq. \eqref{eq:moyal3} for $\eps = 0$. A similar calculation in first order demonstrates that
\bea \label{eq:moyal5}
\cO_1 & = & \frac{i}{2} \nabla_x \mathfrak{a}^{(0)} \cdot \nabla_\eta W^{(0)}- \frac{i}{2} \nabla_\eta \mathfrak{a}^{(0)} \cdot \nabla_x W^{(0)}.
\eea
Higher order contributions can be obtained in a similar fashion but will not be discussed here.

The same calculation can be carried out for the other term of the commutator and one obtains
\bea \label{eq:moyal6}
\cW_\eps \left [ \left (\hat \rho \hat A  \right )(x,y)\right ]& = & W^{(0)} \mathfrak{a}^{(0)}  + \frac{i \eps }{2} \left [\nabla_\eta W^{(0)} \cdot \nabla_x \mathfrak{a}^{(0)} \right. \notag \\
&& \left. - \nabla_x W^{(0)} \cdot \nabla_\eta \mathfrak{a}^{(0)} \right ] + \cO(\eps^2).
\eea
Combining Eqs. \eqref{eq:moyal4}, \eqref{eq:moyal5} and \eqref{eq:moyal6} finally gives the desired result, Eq. \eqref{eq:moyal1},
\bea
\cW_\eps \left [ \cL \left ( A \right )(x,y)\right ] & = &  \left [ \mathfrak{a}^{(0)}, W^{(0)}  \right ] + \frac{i \eps }{2} \left [ \left \{ \mathfrak{a}^{(0)}, W^{(0)} \right \}_{x,\eta} - \left \{ W^{(0)},  \mathfrak{a}^{(0)} \right \}_{x,\eta} \right ]  + \cO(\eps^2),
\eea
where $\{ \cdot , \cdot \}_{x,\eta}$ denotes Poisson's bracket.

\section{Derivation of Eqs. \eqref{eq:q01}} \label{app:dissilimit1}

We rescale Eq. \eqref{eq:quantcoll} in order to obtain
\bea \label{eq:scaledissi}
\frac{1}{\eps} \cQ_\eps (W^{(\eps)})(x, \eta,t) & = & \frac{2(2 \pi \eps)^{2d}}{\eps} \int \dint{\xi} \dint{\xi'} \tilde \kappa^{(\eps)}_{\beta \beta' \alpha \alpha'}(\xi, \xi') \notag \\
&& \times \left [ \tilde V_{\beta \alpha} \left ( \xi \right ) W^{(\eps)} \left (x, \eta - \frac{\xi - \xi'}{2} ,t \right ) \tilde V_{\beta' \alpha'} \left (\xi' \right )  \right. \notag \\
&& - \frac{1}{2} \tilde V_{\beta \alpha} \left ( \xi \right ) \tilde V_{\beta' \alpha'} \left ( \xi' \right ) W^{(\eps)} \left (x, \eta - \frac{\xi + \xi'}{2},t \right ) \notag \\
&& \left. -\frac{1}{2} W^{(\eps)} \left (x, \eta + \frac{\xi + \xi'}{2},t \right ) \tilde V_{\beta \alpha} \left ( \xi \right )V_{\beta' \alpha'} \left ( \xi' \right ) \right ] \notag \\
&& \times \exp \left [\frac{i}{\eps} x \cdot (\xi + \xi' ) \right ].
\eea

According to Eq. \eqref{eq:vinv}, where we used that $\tilde{V}_{\beta \alpha}(\xi)$ is independent of $\eps$ since $\mathbf{V}^{(\eps)}(r)$ is strongly varying in position space. Furthermore, from the definition of $\mathbf{K}$, Eq. \eqref{eq:kappadef}, we obtain that
\bea \label{eq:kappascale}
\mathbf{K}^{(\eps)}(z,z') & = & \gamma \mathbf{C}^{(\eps)}(z - z')+ b \mathbf{D}(z,z'),
\eea
where we already inserted assumption \eqref{eq:scalec1} and employed that due to Eq. \eqref{eq:assum3} $\mathbf{D}^{(\eps)}(z,z') = b \mathbf{D}(z,z')$. Hence, with the help of Eq. \eqref{eq:kinv} we obtain for the matrix elements of Eq. \eqref{eq:kappascale}
\bea
\tilde{\kappa}^{(\eps)}( \xi,\xi') & = & \gamma \tilde c^{(\eps)}_{\alpha \alpha'\beta \beta'}(\xi) \delta(\xi + \xi')
 + \eps \tilde{d}^{(\eps)}_{\alpha \alpha' \beta \beta'}\left ( \xi, \xi' \right ).
\eea
We shall now discuss these two contributions to the integral in Eq. \eqref{eq:scaledissi} separately. We note that the Fourier transform of the strongly varying part of $\mathbf{C}^{(\eps)}$, see Eq. \eqref{eq:scalec1}, is independent of $\eps$, i.e. $\tilde{\mathbf{\Gamma}}_0^{(\eps)} = \tilde{\mathbf{\Gamma}}_0$. Hence, if $\gamma = \eps$ the strongly varying part results in a collision integral of the form \eqref{eq:q01}, for $\gamma = \eps^2$ or even weaker, this contribution vanishes and for $\gamma = 1$ or stronger, the collision integral diverges. In a similar fashion, for the slowly varying part of $\mathbf{C}^{(\eps)}$ we have
\be
\tilde{\mathbf{\Gamma}}^{(\eps)}_s(\xi) = \frac{1}{(2 \pi \eps)^d} \int \dint{z} \mathbf{\Gamma}_s(z) \exp \left ( - \frac{i}{\eps} \xi \cdot z \right ) \equiv \frac{1}{\eps^d} \tilde{\mathbf{\Gamma}}_s \left ( \frac{\xi}{\eps} \right ),
\ee
hence, the resulting collision integral vanishes as $\eps$ approaches zero.


We shall now study the contribution to \eqref{eq:scaledissi} arising from $\mathbf{D}$. Applying the Fourier transform \eqref{eq:kinv} to $\mathbf{D}$ [see Eq. \eqref{replace1}] gives the matrix elements
\bea
\tilde {d}^{(\eps)}_{\alpha \alpha' \beta \beta'}(\xi,\xi') & = & \frac{\delta_{\alpha \alpha'} \delta_{\beta \beta'}}{(2 \pi \eps)^{2d}} \int \dint{z} n_{\alpha \beta}(z) \exp \left [ - \frac{i}{\eps} z \cdot ( \xi + \xi' ) \right ]  \notag \\
 & = & \frac{\delta_{\alpha \alpha'} \delta_{\beta \beta'}}{(2 \pi \eps)^{2d}} \tilde{n}_{\alpha \beta}\left ( \frac{\xi + \xi'}{\eps} \right ).
\eea
Inserting this expression into Eq. \eqref{eq:scaledissi} yields that this contribution tends to zero after the substitution $\xi \to \eps \xi$ and $\\xi' \to \eps \xi'$. The final form of Eq. \eqref{eq:q01} is obtained by performing the steps outlined in App. \ref{app:rewrite}.


\section{Derivation of Eqs. \eqref{eq:qs}} \label{app:dissilimit2}

In order to derive this result we rewrite Eq. \eqref{eq:dissiwig} in rescaled variables as
\bea
\cQ_\eps(W^{(\eps)}) &  : = & \frac{2}{(2 \pi \eps)^d}\int \dint{x'} \dint{z} \dint{z'} \int \dint{\eta'} \kappa^{(\eps)}_{\beta \beta' \alpha \alpha'}(z, z') \notag \\
&& \times \left [ V_{\alpha \beta} \left ( x + \frac{1}{2} x' - z \right ) W^{(\eps)'} V_{\alpha'\beta' } \left ( x - \frac{1}{2} x' - z' \right )  \right. \notag \\
&& - \frac{1}{2} V_{\alpha\beta } \left ( x + \frac{1}{2} x' - z \right ) V_{\alpha'\beta' } \left ( x + \frac{1}{2} x' - z' \right ) W^{(\eps)'} \notag \\
&& \left. -\frac{1}{2} W^{(\eps)'} V_{\alpha\beta } \left ( x - \frac{1}{2} x' - z \right )V_{\alpha'\beta' } \left ( x - \frac{1}{2} x' - z' \right ) \right ] \notag \\
&& \times \exp \left [- \frac{i}{\eps} x' \cdot ( \eta - \eta') \right ],
\eea
where $W^{(\eps)'} = W^{(\eps)}(x,\eta',t)$. We note that $V_{\alpha \beta}(r)$ is independent of $\eps$ since $\mathbf{V}$ is slowly varying in position space.

We now study the different scenarios arising from Eq. \eqref{eq:scalec2}. For the slowly varying part of $\mathbf{C}^{(\eps)}$ we substitute $x' \to \eps x'$, draw the limit and integrate with respect to $x'$ and $\eta'$ in order to obtain Eq. \eqref{eq:qs1} for $\gamma = \eps$ in the low density limit and for $\gamma =  1 /\eps$ in the weak coupling limit.

For the strongly varying part $\mathbf{C}^{(\eps)}$ we also substitute $z \to \eps z$ and $z' \to \eps z'$. Hence, $z$ and $z'$ only appear in the matrix elements of $\mathbf{\Gamma}_0$ as $\eps \to 0$. However, according to Eq. \eqref{eq:defc} this integral vanishes at all scales, thus, no contribution arises from the strongly varying part $\mathbf{\Gamma}_0^{(\eps)}$ in the semiclassical limit. Finally, in the low density limit the contribution of $\mathbf{D}^{(\eps)}(z,z') = \eps \mathbf{D}(z,z')$ is easily seen to be of the form \eqref{eq:qn}. We remark this term vanishes in the weak-coupling limit.


\begin{thebibliography}{10}
\bibitem{fabian1}
I.~Zutic, J.~Fabian, and Das~Sarma S.
\newblock Spintronics: Fundamentals and applications.
\newblock {\em Rev. Mod. Phys.}, 76(2):323--410, 2004.

\bibitem{fabian2}
J.~Fabian, A.~Matos-Abiague, Ertler C., P.~Stano, and I.~Zutic.
\newblock {\em Acta Physica Slovaca}, 57:565 -- 907, 2007.

\bibitem{zhang02}
S.~Zhang, P.M. Levy, and A.~Fert.
\newblock Mechanisms of spin-polarized current-driven magnetization switching.
\newblock {\em Phys. Rev. Lett.}, 88(23):236601--1, 2002.

\bibitem{garcia06}
C.J. Garcia-Cervera and X.-P. Wang.
\newblock Spin-polarized currents in ferromagnetic multilayers.
\newblock {\em J. Comp. Phys.}, 224:699, 2007.

\bibitem{possanner10}
S.K. Possanner and N.~Ben~Abdallah.
\newblock Spin-transfer torques: Self-consistent solution of the spin-diffusion
  equation and the landau-lifshitz equation.
\newblock pages 37--40, Sept.

\bibitem{simanek}
E.~Simanek.
\newblock Spin accumulation and resistance due to a domain wall.
\newblock {\em Phys. Rev. B}, 63:224412, 2001.

\bibitem{raymond}
R.~El~Hajj.
\newblock {\em Etude math\'{e}matique et num�rique de mod\`{e}les de
  transport: application \`{a} la spintronique}.
\newblock PhD thesis, Institut de Math\'{e}matiques de Toulouse (IMT),
  Universit\'{e} Paul Sabatier, 2008.

\bibitem{MeClaudia2011}
S.~Possanner and C.~Negulescu.
\newblock Diffusion limit of a generalized matrix boltzmann equation for
  spin-polarized transport.
\newblock {\em Kinetic and Related Models}, 4(4):1159--1191, December 2011.

\bibitem{nielsen}
Michael Nielsen and Isaac Chuang.
\newblock {\em {Q}uantum {C}omputation and {Q}uantum {I}nformation}.
\newblock {C}ambridge {U}niversity {P}ress, Cambridge, 2000.

\bibitem{awschalom}
D.D. Awschalom, D.~Loss, and N.~Samarth.
\newblock {\em Semiconductor Spintronics and Quantum Computation}.
\newblock Series on Nanoscience and Technology. Springer-Verlag Berlin, 2002.

\bibitem{zeilinger}
K.~Hornberger, S.~Uttenthaler, B.~Brezger, L.~Hackerm{\"u}ller, M.~Arndt, and
  A.~Zeilinger.
\newblock Collisional decoherence observed in matter wave interferometry.
\newblock {\em Physical review letters}, 90(16):160401, 2003.

\bibitem{jauho}
Lino Reggiani, Paolo Lugli, and A.~P. Jauho.
\newblock Quantum kinetic equation for electronic transport in nondegenerate
  semiconductors.
\newblock {\em Phys. Rev. B}, 36:6602--6608, Oct 1987.

\bibitem{spicka}
V\'aclav \ifmmode \check{S}\else \v{S}\fi{}pi\ifmmode~\check{c}\else
  \v{c}\fi{}ka and Pavel Lipavsk\'y.
\newblock Quasiparticle boltzmann equation in semiconductors.
\newblock {\em Phys. Rev. B}, 52:14615--14635, Nov 1995.

\bibitem{hornberger}
Klaus Hornberger and Bassano Vacchini.
\newblock Monitoring derivation of the quantum linear boltzmann equation.
\newblock {\em Phys. Rev. A}, 77:022112, Feb 2008.

\bibitem{fert}
A.~Fert.
\newblock Nobel lecture: Origin, development, and future of spintronics.
\newblock {\em Rev. Mod. Phys.}, 80:1517, 2008.

\bibitem{thiaville}
F.~Pi\'echon and A.~Thiaville.
\newblock Spin transfer torque in continuous textures: Semiclassical boltzmann
  approach.
\newblock {\em Phys. Rev. B}, 75:174414, 2007.

\bibitem{xiao}
J.~Xiao, A.~Zangwill, and M.D. Stiles.
\newblock A numerical method to solve the boltzmann equation for a spin valve.
\newblock {\em Eur. Phys. J. B}, 59:415--427, 2007.

\bibitem{culcer}
Dimitrie Culcer, Jairo Sinova, N.~A. Sinitsyn, T.~Jungwirth, A.~H. MacDonald,
  and Q.~Niu.
\newblock Semiclassical spin transport in spin-orbit-coupled bands.
\newblock {\em Phys. Rev. Lett.}, 93:046602, Jul 2004.

\bibitem{zhang}
J.~Zhang, P.M. Levy, S.~Zhang, and V.~Antropov.
\newblock Identification of transverse spin currents in noncollinear magnetic
  structures.
\newblock {\em Phys. Rev. Lett.}, 93(256602), 2004.

\bibitem{fertscattering}
C.~Vouille, A.~Barth\'el\'emy, F.~Elokan~Mpondo, A.~Fert, P.~A. Schroeder,
  S.~Y. Hsu, A.~Reilly, and R.~Loloee.
\newblock Microscopic mechanisms of giant magnetoresistance.
\newblock {\em Phys. Rev. B}, 60(9):6710, 1999.

\bibitem{viret}
M.~Viret, D.~Vignoles, D.~Cole, J.M.D. Coey, W.~Allen, S.D. Daniel, and J.F.
  Gregg.
\newblock Spin scattering in ferromagnetic thin films.
\newblock {\em Phys. Rev. B}, 53(13):8464, 1996.

\bibitem{zachos}
C.~Zachos, D.~Fairlie, and T.~Curtright.
\newblock {\em Quantum mechanics in phase space: an overview with selected
  papers}, volume~34.
\newblock World Scientific, 2005.

\bibitem{gerard}
Patrick Gérard, Peter~A. Markowich, Norbert~J. Mauser, and Frédéric Poupaud.
\newblock Homogenization limits and wigner transforms.
\newblock {\em Communications on Pure and Applied Mathematics}, 50(4):323--379,
  1997.

\bibitem{jauhobook}
H.~Haug and A.-P. Jauho.
\newblock {\em Quantum kinetics in transport and optics of semiconductors}.
\newblock Springer - Berlin, 1996.

\bibitem{erdos2000}
László Erdős and Horng-Tzer Yau.
\newblock Linear boltzmann equation as the weak coupling limit of a random
  schrödinger equation.
\newblock {\em Communications on Pure and Applied Mathematics}, 53(6):667--735,
  2000.

\bibitem{spohn77}
Herbert Spohn.
\newblock Derivation of the transport equation for electrons moving through
  random impurities.
\newblock {\em Journal of Statistical Physics}, 17:385--412, 1977.

\bibitem{erdos2005}
David Eng and Laszlo Erdos.
\newblock The linear boltzmann equation as the low density limit of a random
  schrÖdinger equation.
\newblock {\em Reviews in Mathematical Physics}, 17(06):669--743, 2005.

\bibitem{pulvirenti}
D.~Benedetto, F.~Castella, R.~Esposito, and M.~Pulvirenti.
\newblock Some considerations on the derivation of the nonlinear quantum
  boltzmann equation.
\newblock {\em Journal of Statistical Physics}, 116:381--410, 2004.

\bibitem{lindblad}
G.~Lindblad.
\newblock On the generators of quantum dynamical semigroups.
\newblock {\em Communications in Mathematical Physics}, 48(2):119--130, 1976.

\bibitem{breuer}
H.P. Breuer and F.~Petruccione.
\newblock {\em The theory of open quantum systems}, volume~28.
\newblock Oxford University Press Oxford, 2002.

\bibitem{rivas}
A.~Rivas and S.~Huelga.
\newblock {\em Open Quantum Systems: An Introduction}.
\newblock Springer Verlag, 2011.

\bibitem{almeida}
A~M~Ozorio de~Almeida, P~de~M~Rios, and O~Brodier.
\newblock Semiclassical evolution of dissipative markovian systems.
\newblock {\em Journal of Physics A: Mathematical and Theoretical},
  42(6):065306, 2009.

\bibitem{possanner12}
S.~Possanner and B.~A. Stickler.
\newblock Non-markovian quantum dynamics from environmental relaxation.
\newblock {\em Phys. Rev. A}, 85:062115, Jun 2012.

\bibitem{kossakowski}
V.~Gorini, A.~Frigerio, M.~Verri, A.~Kossakowski, and ECG Sudarshan.
\newblock Properties of quantum markovian master equations.
\newblock {\em Reports on Mathematical Physics}, 13(2):149--173, 1978.

\bibitem{Note1}
Denoting by $\tau _0$ the characteristic timescale of the system corresponding
  to the Hamiltonian $\protect \mathaccentV {hat}05EH^{mf}_0$, the
  characteristic energy is defined via $\epsilon _0 \tau _0 = \hbar $.

\bibitem{esposito11}
Massimiliano Esposito and Shaul Mukamel.
\newblock Fluctuation theorems for quantum master equations.
\newblock {\em Phys. Rev. E}, 73:046129, Apr 2006.

\bibitem{Note2}
We note that in the case of more than two spin degrees of freedom one may
  express all quantities in a form analogous to Eq. \eqref {eq:paulibase} by
  employing the generalized Gell-Mann matrices\cite {bertlmann08}.

\bibitem{white}
R.M. White.
\newblock {\em Quantum Theory of Magnetism: magnetic properties of materials}.
\newblock Springer, 2007.

\bibitem{ashcroft}
N.W. Ashcroft and N.D. Mermin.
\newblock {\em Solid State Physics}.
\newblock Saunders College Publishing, 1976.

\bibitem{kenfack}
Anatole Kenfack and Karol Życzkowski.
\newblock Negativity of the wigner function as an indicator of
  non-classicality.
\newblock {\em Journal of Optics B: Quantum and Semiclassical Optics},
  6(10):396, 2004.

\bibitem{cercignani}
C.~Cercignani.
\newblock {\em The Boltzmann equation and its applications}, volume~67.
\newblock Springer, 1988.

\bibitem{Note3}
This statement is is easily proved: suppose ${\protect \mathscr D}(\Sigma
  ^\dagger W' \Sigma ) \not =0$. We may then define a matrix $A = \Sigma
  {\protect \mathscr D}(\Sigma ^\dagger W' \Sigma ) \Sigma ^\dagger \in
  \protect \mathrm {Ker}({\protect \mathcal Q}_\hbar ^{(2)})$, which is a
  contradiction to the assumption $W' \in \protect \mathrm {Ker}({\protect
  \mathcal Q}_\hbar ^{(2)})^\bot $, which proofs '$\Leftarrow $'. On the other
  hand, let $W' \in \protect \mathrm {Ker}({\protect \mathcal Q}_\hbar
  ^{(2)})^\bot $ and $B \in \protect \mathrm {Ker}({\protect \mathcal Q}_\hbar
  ^{(2)})$. Then we have \begin {equation}0 = \protect \mathrm {tr}\left (W'
  B\right ) = \protect \mathrm {tr}\left (\Sigma ^\dagger W'\Sigma \Sigma
  ^\dagger B \Sigma \right ) = \protect \mathrm {tr}\left (\Sigma ^\dagger B
  \Sigma {\protect \mathscr D}(\Sigma ^\dagger B \Sigma ) \right ) \Rightarrow
  \Sigma ^\dagger W' \Sigma = {\protect \mathscr O}(\Sigma ^\dagger W' \Sigma
  ), \end {equation}which demonstrates '$\Rightarrow $' and the proof is
  completed.

\bibitem{Note4}
It is interesting to note that the covariance in the collision integral is a
  direct consequence of the requirement of well-posedness of the hierarchy of
  master equations on the quantum scale\cite {possanner12}.

\bibitem{morandi}
O.~Morandi and F.~Sch\"{u}rrer.
\newblock Wigner model for quantum transport in graphene.
\newblock {\em Journal of Physics A: Mathematical and Theoretical},
  44(26):265301, 2011.

\bibitem{bertlmann08}
Reinhold~A Bertlmann and Philipp Krammer.
\newblock Bloch vectors for qudits.
\newblock {\em Journal of Physics A: Mathematical and Theoretical},
  41(23):235303, 2008.

\end{thebibliography}
\end{document}